\documentclass[%
 reprint,
 amsmath,amssymb,
 aps,
]{revtex4-2}

\usepackage{graphicx}
\usepackage{dcolumn}
\usepackage{bm}
\usepackage{siunitx} 
\usepackage{makecell} 
\usepackage{gensymb}

\makeatletter
\def\blfootnote{\xdef\@thefnmark{}\@footnotetext}
\makeatother

\usepackage[
text={7.3in,9.1in},
]{geometry}

\begin{document}

\title{Size-dependent self-avoidance enables superdiffusive migration in macroscopic unicellulars}

\author{Lucas Tr\"oger}
\author{Florian Goirand}
\author{Karen Alim}%
 \email{k.alim@tum.de}
\affiliation{%
 Technical University of Munich,  TUM School of Natural Sciences, Department of Bioscience, \\ Center for Protein Assemblies (CPA),  85748 Garching, Germany
}%

\begin{abstract}
Many cells face search problems, such as finding food, mates or shelter, where their success depends on their search strategy. In contrast to other unicellular organisms, the slime mold \textit{Physarum polycephalum} forms a giant network-shaped plasmodium while foraging for food. What is the advantage of the giant cell on the verge of multicellularity? We experimentally study and quantify the migration behavior of \textit{P.~polycephalum} plasmodia on the time scale of days in the absence and presence of food. We develop a model which successfully describes its migration in terms of ten data-derived parameters. Using the mechanistic insights provided by our data-driven model, we find that regardless of the absence or presence of food, \textit{P.~polycephalum} achieves superdiffusive migration by performing a self-avoiding run-and-tumble movement. In the presence of food, the run duration statistics change, only controlling the short-term migration dynamics. However, varying organism size, we find that the long-term superdiffusion arises from self-avoidance determined by cell size, highlighting the potential evolutionary advantage that this macroscopically large cell may have.
\end{abstract}

\maketitle


\section{\label{sec:level1} Introduction}

Search problems are a widespread challenge for living organisms, from unicellular species to animals \cite{berg1972chemotaxis, li2008persistent,jackson2006communication,sims2008scaling,raichlen2014evidence}. Success in finding food, mates or shelter depends on the search strategy \cite{bartumeus2005animal,shaebani2022distinct,palyulin2014levy,faustino2007search,rupprecht2016optimal}. The search of unicellulars is commonly characterized by Brownian motion on long time scales \cite{berg1972chemotaxis, li2008persistent, furth1920brownsche, wu2014three}, whereas more evolved life forms such as mammalian cells, insects, birds or humans show superdiffusion \cite{dieterich2008anomalous,harris2012generalized,huda2018levy,upadhyaya2001anomalous,miramontes2014levy,viswanathan1996levy,vilk2022ergodicity, raichlen2014evidence}. Strikingly, unicellulars in groups, like swarming bacteria \cite{ariel2015swarming}, and unicellular organisms on the verge of multicellularity, like large multinucleate cells \cite{rodiek2015migratory,shirakawa2019biased}, also show superdiffusion. Since search success is coupled to survival chances \cite{faustino2007search}, the need to find effective strategies could be a driver of evolution. With the increase in cell size being the putative first evolutionary step in the critical transition from unicellular to multicellular life \cite{bonner1998origins}, what are the benefits of a large cell size in the context of search problems?

The mean squared displacement (MSD) measures the deviation of an organism from a reference position during its migration, quantifying how much space is explored. The MSD is typically characterized by a power law $\mathrm{MSD}(t) \propto D t^{\beta}$ with generalized diffusion constant $D$, time $t$ and exponent $\beta$ \cite{metzler2014anomalous}. Scale-invariant migration trajectories allow for a constant exponent to classify the MSD, bounded by $\beta=1$ for Brownian motion and $\beta=2$ for ballistic motion, where values between $1$ and $2$ refer to superdiffusion, as arising in L\'evy walks and self-avoiding walks. For L\'evy walks, the exponent is determined by the power law describing the step length distribution  \cite{viswanathan2008levy}, for a self-avoiding walk in two dimensions the exponent is generally accepted to be close to $\beta=1.5$ \cite{slade1994self,slade2019self}. Depending on the search task, different migration strategies lead to different success, with Brownian motion being best when search targets are close \cite{palyulin2014levy}, but superdiffusion outperforming it when target density is low \cite{faustino2007search}. Ballistic motion is predicted to be optimal for Poisson-distributed targets but is less efficient than a run-and-tumble walk when the goal is to find a fixed target within a confinement \cite{rupprecht2016optimal}. Unicellulars already have mixed search strategies, performing ballistic motion on short time scales and Brownian motion on long time scales. Prominent examples include run-and-tumble motion and persistent random walks \cite{berg1972chemotaxis, li2008persistent, li2011dicty, furth1920brownsche, wu2014three}, which are characterized by a transition from ballistic motion on short time scales to Brownian motion on long time scales \cite{furth1920brownsche,bartumeus2005animal,shaebani2019transient}. Superdiffusion on long time scales arises, for example, from long-range correlations as found for higher life forms \cite{bartumeus2005animal,dieterich2008anomalous}, but also in crowded bacterial swarms \cite{ariel2015swarming} or in the giant cells of \textit{Physarum polycephalum} \cite{rodiek2015migratory,shirakawa2019biased}. Yet, how the giant unicellular \textit{P.~polycephalum} generates superdiffusive motion is unclear despite its fascinating life form on the verge of multicellularity \cite{bonner1998origins}. 

\begin{figure*}[t]
\centering
\includegraphics[width=\textwidth]{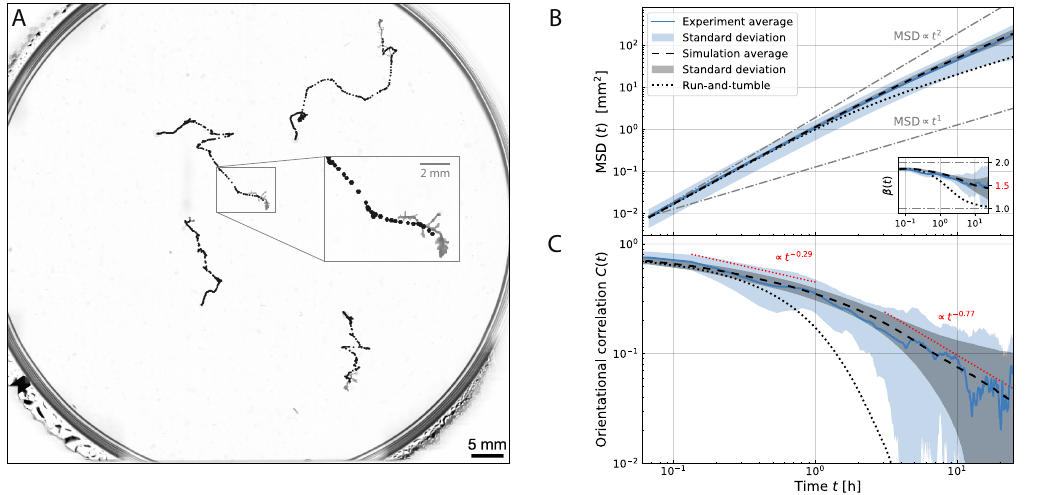}
\caption{Migration of \textit{P.~polycephalum} shows superdiffusion and anomalous persistence. (A) Plasmodia of \textit{P.~polycephalum} on a petri dish with 1.5\% agar (background subtracted image) with overlayed trajectories (one dot every $\SI{8}{min}$, \SI{40}{h} in total). (B) Log-log plot of the superdiffusive mean squared displacement (MSD) of migrating plasmodia. Blue line and shaded region represent the ensemble average over the time-averaged MSDs of 14 individual trajectories and the standard deviation, respectively. The black dashed line and shaded region shows corresponding simulation results. The black dotted line shows results from simulations without the self-avoidance, so a pure run-and-tumble. Dash-dotted lines show the MSDs of ballistic ($\propto t^2$) and diffusive motion ($\propto t^1$). Inset: Instantaneous MSD exponent $\beta(t)$. The migration is superdiffusive ($\beta > 1$) over $\approx3$ orders of magnitude in time. Red tick: Flory exponent for a self-avoiding walk in two dimensions. (C) Log-log plot of the orientational correlation $C(t)$ as a measure of persistence. The red dotted lines show power laws for comparison.} 
\label{fig:setup}
\end{figure*}

\textit{P.~polycephalum} is a plasmodial slime mold -- a unicellular, non-dividing, multinucleate organism, devoid of the complexity of multicellular model systems. Despite the absence of a nervous system, \textit{P.~polycephalum} coordinates complex behaviors including adaptive network formation \cite{tero2010rules}, speed-accuracy trade-offs during foraging \cite{latty2011speed}, and nutritional decisions \cite{dussutour2010amoeboid}, which has earned it a reputation for being ``smart'' \cite{verge2023physarum}. It can also escape traps by leaving a trail of slime, which it uses as an external memory to actively avoid areas it has already visited \cite{reid2012slime, reid2013amoeboid, smith2017hansel}. Its cell size is highly variable, covering a range of orders of magnitude from $\SI{100}{\micro\meter}$ to meters \cite{verge2023physarum}, and correlates with the average speed of locomotion, which is oscillating periodically \cite{kuroda2015allometry}. 
\textit{P.~polycephalum} has proven to be highly amenable to both observation and quantification of its dynamics, as well as theoretical modeling, making it an ideal model system to, here, develop mechanistic insight into its superdiffusive motion.
\newpage
 In this work, we experimentally observe \textit{P.~polycephalum} plasmodia migrating in a neutral environment, perform data analysis to determine their migration characteristics and develop a data-driven model which captures \textit{P.~polycephalum}'s migration behavior. We show the robustness of our model by quantifying how the behavior changes in a nutritious migration environment. Varying organism size, we reveal the pivotal role of cell size in driving superdiffusive motion by enabling reliable self-avoidance only above a cell size of $\SI{0.65}{mm^2}$. Our results highlight the adaptive capabilities of \textit{P.~polycephalum}, as well as the impact of cell size on space exploration performance, suggesting the potential evolutionary advantage that this large unicellular may have.

\begin{figure*}[t]
\centering
\includegraphics[width=\textwidth]{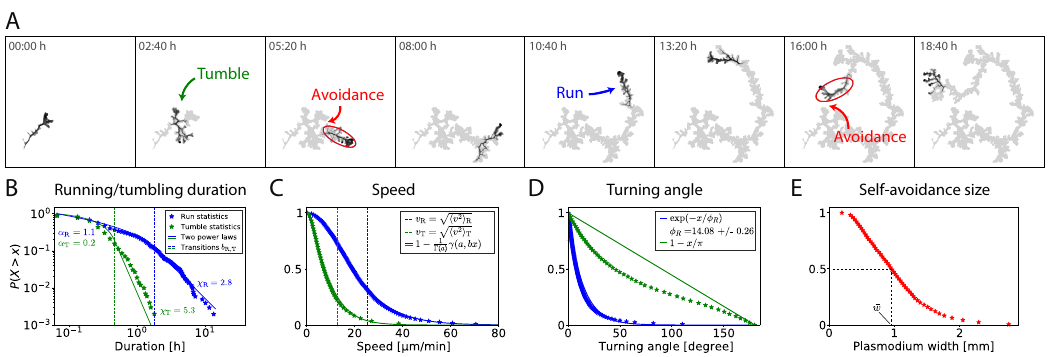}
\caption{\textit{P.~polycephalum} performs a self-avoiding run-tumble movement during its migration. (A) Exemplary trajectory of a plasmodium with highlighted events of tumbling (green arrow), running (blue arrow) and self-avoidance (red arrows). The cumulative visited area is visualized in light gray. Overlayed ellipses are fitted to the plasmodium as an approximation for its size. Scale bar: 2 mm. (B)-(E) Parameter extraction from data for the model: Complementary cumulative distribution functions (CCDF), $P(X>x)$, of the analyzed variables, with $X$ denoting the respective variable. Fits from maximum likelihood estimation in solid lines. (B) Run and tumble durations fitted with a combination of two power laws. (C) Speeds during runs and tumbles with fitted gamma distributions. Run statistics: $a=2.60, b=\SI{0.12}{\mathrm{min}^{-1}}$. Tumble statistics: $a=1.69,\,b=\SI{0.18}{\mathrm{min}^{-1}}$. Dashed lines: Root mean squared speeds. (D) Turning angles between consecutive steps during runs and tumbles. Run statistics fitted by an exponential distribution and tumble statistics approximated by a homogeneous distribution. (E) Widths of the plasmodia slime trails as a measure for the avoided space around the trajectories. The width is estimated as the width of an ellipse fitted to the plasmodium as in (A). The dashed line shows the median width of all plasmodia.}
\label{fig:parameters}
\end{figure*}

\section{Results}
\subsection{Migration shows superdiffusive MSD and anomalous persistence}
We experimentally follow and quantify the migration of \textit{P.~polycephalum} plasmodia of different size and on different migration substrates. In the initial setup, plasmodia are allowed to migrate on a two-dimensional, non-nutritious substrate (1.5\% agar). Using bright-field microscopy in combination with a stage-top incubator, we follow the movement of plasmodia over a period of up to 134 hours (5.6 days) while ensuring constant and homogeneous environmental conditions, namely humidity, light and temperature. We perform cell tracking and statistically analyze the centroid trajectories of individual plasmodia (Fig.~\ref{fig:setup}).

To quantify the space exploration behavior of \textit{P.~polycephalum}, we use two established measures: the time-averaged mean squared displacement, MSD,
\begin{equation*}
    \mathrm{MSD}(t) = \langle\left[\mathbf{r}(\tau+t)-\mathbf{r}(\tau)\right]^2\rangle^{}_\tau,
\end{equation*}
and the orientational correlation, C,
\begin{equation*}
    C(t) = \langle\cos\theta\rangle=\langle\hat{\mathbf{r}}(\tau+t)\cdot\hat{\mathbf{r}}(\tau)\rangle_\tau,
\end{equation*}
where $\mathbf{r}(\tau)$ is the position of the plasmodium's centroid at time $\tau$ with corresponding unit vector $\hat{\mathbf{r}}(\tau)$,  and $t$ is the time interval of the MSD measurement.
 How does the MSD of migrating \textit{P.~polycephalum} scale with $t$? Our statistical analysis, based on 14 experimental trajectories, shows that the MSD of plasmodia migrating on plain agar is superdiffusive over a large range of time scales (Fig.~\ref{fig:setup}B). However, we find that the MSD exponent $\beta$ is not constant, but depends on the time scale $t$. Defining the instantaneous MSD exponent $\beta(t)$ as the logarithmic derivative of the MSD
\begin{equation*}
    \beta(t) = \frac{\partial\log\mathrm{MSD}(t)}{\partial\log{t}},
\end{equation*}
we find that $\beta(t)$ is decreasing towards longer time scales  (Fig.~\ref{fig:setup}B, inset): at the smallest time scale ($\SI{4}{min}$), $\beta$ is close to a value of 2, signifying almost ballistic motion. At larger times, $\beta$ is slowly decaying, approaching a seemingly stable value of $\beta\approx1.5$ at the order of 10 hours, which is compatible with a self-avoiding walk.
In addition to the MSD, the orientational correlation describes how well the direction of migration is aligned with respect to the direction at previous time points, thus quantifying the persistence of the migration. In the case of \textit{P.~polycephalum}, we observe that orientational correlations show a decay slower than exponential (Fig.~\ref{fig:setup}C), which contrasts with the exponential decay for persistent random walks \cite{viswanathan2005necessary,li2011dicty}. Thus, \textit{P.~polycephalum}'s migration shows anomalous persistence.
Taken together, both experimentally observed MSD and orientational correlation cannot be explained by the standard random walk models discussed in the introduction. To understand how superdiffusivity and anomalous persistence emerge, we next quantify the migration characteristics.

The trajectories of \textit{P.~polycephalum} are reminiscent of a run-and-tumble motion \cite{rodiek2015patterns} with phases of straight movement and phases of stationarity, after which plasmodia change their direction (Fig.~\ref{fig:setup}A). The stretches of straight movement are consistent with an MSD exponent close to $\beta=2$ on small time scales, associated with ballistic migration. Another important characteristic of \textit{P.~polycephalum}'s migration is that plasmodia usually do not cross their own path. This behavior is a result of their path-marking mechanism: migrating plasmodia leave behind a slime trail which they generally avoid when encountered \cite{reid2012slime, reid2013amoeboid, smith2017hansel}. Although previously considered irrelevant in the case of very small, tadpole-shaped plasmodia \cite{rodiek2015migratory}, this avoidance could explain \textit{P.~polycephalum}'s superdiffusive behavior since self-avoiding walks are associated with a superdiffusive MSD exponent \cite{slade1994self,slade2019self}. 
Taken together, these observations suggest that the migration of \textit{P.~polycephalum} could be described by a combination of two types of random walks, i.e.~as a self-avoiding run-and-tumble walk. To investigate this hypothesis, we next quantify the statistics of the migration behavior.

\begin{table*}[t]
\caption{Model parameters related to the migration rules extracted from the data and their control over the MSD.}
\begin{ruledtabular}
\begin{tabular}[t]{lcl}
Migration rule extracted from data & Model parameters & Control over MSD\\
\hline
\makecell[tl]{Run and tumble durations\\ distributed as two power laws}      &       &  \makecell[tl]{Generalized diffusion constant, \\MSD exponent at small \\time scales ($\lesssim\SI{1}{h}$)}\\
\hspace{0.5cm} - exponent of first power law   & $\alpha_\mathrm{R,T}$      &\\
\hspace{0.5cm} - exponent of second power law   & $\chi_\mathrm{R,T}$      &\\
\hspace{0.5cm} - transition point between the power laws   & $b_\mathrm{R,T}$      &\\
Constant speed during run and tumble                                    & $v_\mathrm{R,T}$                  & Generalized diffusion constant\\
Turning angles distributed exponentially during runs   & $\phi_\mathrm{R}$                             & \makecell[tl]{MSD exponent at short and in-\\termediate time scales ($\lesssim\SI{10}{h}$)}\\
Self-avoidance according to plasmodium width                             & $\bar{w}$                                           & \makecell[tl]{MSD exponent at intermediate \\and long time scales ($\gtrsim\SI{1}{\hour}$)}\\
\end{tabular}
\end{ruledtabular}
\label{table:parameters}
\end{table*}%

\subsection{Quantifying Run-and-Tumble and Self-Avoiding Behavior}
Closely inspecting our experimental trajectories, we distinguish \textit{P.~polycephalum}'s phases of fast, straight movement and phases of stationarity with small positional oscillations, similar to run-and-tumble \cite{berg1972chemotaxis} or intermittent behavior \cite{benichou2011intermittent}. We discriminate the two phases, i.e.~run and tumble, by measuring the local directionality, or straightness, of the movement. This quantity is defined as the ratio of the distance between two positions and the length of the plasmodium's path connecting these two positions \cite{varennes2019physical}. Sections of the trajectory with a directionality greater than a threshold (0.9, Materials and Methods) are identified as phases of running. Phases of arrested movement (tumbling) are characterized by a lower directionality. This method allows us to divide the trajectories into a series of two alternating phases (\textit{SI Appendix}, Fig.~S1), which we analyze separately. We describe a single phase of running or tumbling as a series of steps, characterized by three parameters: phase duration, speeds and turning angles. We find that the distributions of running and tumbling durations are best fitted by combinations of two power laws (Fig.~\ref{fig:parameters}B and \textit{SI Appendix} Fig.~S2A), according to the Akaike information criterion (\textit{SI Appendix}, supporting text). Speed distributions are well described by gamma distributions (Fig.~\ref{fig:parameters}C). The turning-angle distribution during runs is well fitted by an exponential and the one during tumbles is close to a uniform distribution (Fig.~\ref{fig:parameters}D).

In accordance with the literature, we observe that plasmodia generally avoid previously visited areas marked by their slime trail \cite{reid2012slime, reid2013amoeboid, smith2017hansel}. The shape and extent of the slime trail is determined by the cell shape of the migrating plasmodia, which is highly variable in time. We describe the shape of the just forming slime trail by an ellipse fitted to the plasmodium (Fig.~\ref{fig:parameters}A), and approximate the trail width as the median width of the fitted ellipse (Fig.~\ref{fig:parameters}E). In our experimental observations, the distance from which the plasmodia detect the slime trail is often greater than zero (see Fig.~\ref{fig:parameters}A, red arrows), probably due to diffusion of the unknown repellent slime component in the agar. Reid et al.~found that slime from a large culture of slime mold induced an avoidance response in very large plasmodia ($\SI{1}{\cm^2}$ cell area) for up to 6 days \cite{reid2013amoeboid}, which is much longer than the time frame considered in our analysis.

Following the quantification of migration trajectories in terms of distributions, we extract from our data that plasmodia perform a self-avoiding run-and-tumble movement governed by the following rules: a migrating plasmodium alternates between two phases of movement, running and tumbling. Running and tumbling durations are distributed according to a combination of two power laws. In each phase, the plasmodium moves with a speed characteristic for this phase, distributed according to gamma distributions. After each step, it chooses the direction of the next step according to an exponential distribution for runs and according to a uniform distribution for tumbles. As it explores its environment, the plasmodium does not come closer to its past trajectory than the median width of a plasmodium.

\subsection{Model Reveals Individual Contributions of Parameters}
We now use the data-derived rules and statistics to model \textit{P.~polycephalum}'s migration behavior as the combination of run-and-tumble and self-avoidance, with the aim of reproducing \textit{P.~polycephalum}'s movement in terms of the experimentally observed MSD and orientational correlation. We simplify the model by assigning constant speeds during runs and tumbles, rather than sampling from the gamma distributions. With this, the number of parameters in our model reduces to ten: the exponents $\alpha_\mathrm{R,T}$ and $\chi_\mathrm{R,T}$, and the transition points $b_\mathrm{R,T}$ of the combined power law distributions of running and tumbling durations, the average speeds during runs and tumbles $v_\mathrm{R,T}$ corresponding to the root mean squared speeds $\sqrt{\langle v_\mathrm{R,T}^2 \rangle}$, the average turning angle during runs $\phi_\mathrm{R}$ and the median width of the plasmodia $\bar{w}$, which we all extract directly from the data without the need for model fitting. This is done by estimating all parameters by fitting the distributions to the data (Fig.~\ref{fig:parameters}B-E).

\begin{figure*}
\centering
\includegraphics[width=\textwidth]{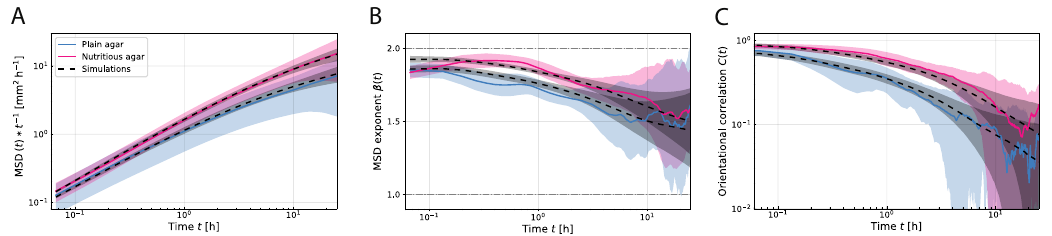}
\caption{MSD exponent and orientational correlation of \textit{P.~polycephalum} are higher on a nutritious migration substrate, but are captured by the same type of model. (A) Log-log plot of the MSD divided by time  of migrating plasmodia on  non-nutritious substrate (14 trajectories) and on nutritious substrate (9 trajectories) and the MSD of the simulated trajectories. Shaded regions show the standard deviations. (B) Log-lin plot of the instantaneous MSD exponent $\beta$ in experiments and simulations. (C) Log-log plot of the orientational correlation $C$ in experiments and simulations.}
\label{fig:nutritious}
\end{figure*}

Now using the ten data-derived parameters, we simulate self-avoiding trajectories as a series of runs and tumbles, which themselves consist of series of random steps characterized by a direction and by a step length, according to the described distributions. In this sense, our simulation is a kinetic process as the trajectories are built step by step in time, which differs from the classical definition of self-avoiding walks in the context of equilibrium polymers. However, the kinetic definition is believed to be in the same universality class as the classical self-avoiding walk and to have the same critical exponents \cite{hemmer1986trapping}. We quantify the MSD and the orientational correlation of the simulated trajectories and find that they reproduce the experimental results with excellent accuracy (Fig.~\ref{fig:setup}B,C), confirming that our model captures the migration behavior of \textit{P.~polycephalum} plasmodia. With the model at hand, we can vary the parameters to understand their individual contributions to the space exploration behavior (see \textit{SI Appendix}, Fig.~S3 for parameter sweeps). The MSD is fully characterized by the generalized diffusion constant $D$ and the time-dependent MSD exponent $\beta(t)$. We find that $D$ depends on the speeds $v_\mathrm{R,T}$ but also on the average running and tumbling durations $\langle R\rangle$ and $\langle T\rangle$, which are determined by $\alpha_\mathrm{R,T}$, $\chi_\mathrm{R,T}$ and $b_\mathrm{R,T}$ (\textit{SI Appendix}, supporting text). This dependence of $D$ can also be seen analytically: we find $D \propto \langle R\rangle\langle v_\mathrm{R}^2\rangle+\langle T\rangle\langle v_\mathrm{T}^2\rangle$ (\textit{SI Appendix}, supporting text). For the MSD exponent $\beta(t)$, we find that the parameters act at different time scales. The MSD exponent at short time scales ($\lesssim\SI{1}{h}$) is determined by $\langle R\rangle$ and $\langle T\rangle$  and the average turning angle during runs $\phi_\mathrm{R}$, with $\beta$ correlating positively with $\langle R\rangle$ and negatively with $\langle T\rangle$ and $\phi_\mathrm{R}$ (\textit{SI Appendix}, Fig.~S3A-C). The MSD exponent at intermediate time scales ($\SI{1}{h}-\SI{10}{h}$) is determined by $\phi_\mathrm{R}$ and the width $\bar{w}$ of the plasmodia, with $\beta$ correlating again negatively with $\phi_\mathrm{R}$ and positively with $\bar{w}$ (\textit{SI Appendix}, Fig.~S3C,D). The MSD exponent at long time scales ($\gtrsim\SI{10}{h}$) is only determined by $\bar{w}$, which controls how fast $\beta$ is converging eventually to the expected value of 1.5 for very long time scales (\textit{SI Appendix}, Figs.~S3 and S4). Table \ref{table:parameters} summarizes these results. The intuition behind the findings is that the straighter the movement, the higher the MSD exponent. This is the case for longer runs and a smaller average turning angle, but also for a larger plasmodium width; the width $\bar{w}$ determines the self-avoidance range, with a larger width inducing more avoidance, resulting in straighter movement. The influence of the parameters on the orientational correlation is similar to the one on the MSD exponent.

Apart from the success of the model and the mechanistic insight it provides, it is not clear how robustly it describes any \textit{P.~polycephalum} plasmodium irrespective of its environment. Also, which migration parameters would change in a different environment?

\begin{figure*}[t]
\centering
\includegraphics[width=\textwidth]{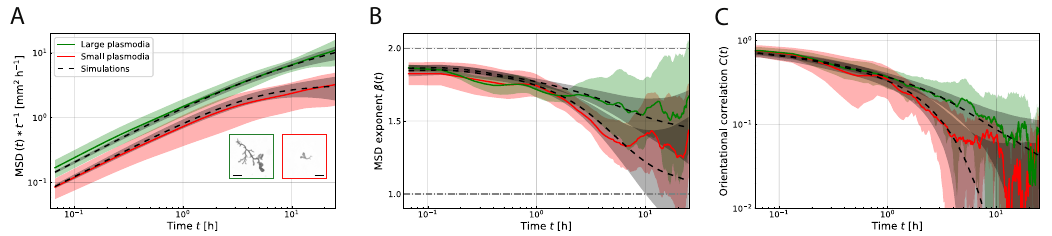}
\caption{MSD exponent and orienational correlation of small plasmodia decrease strongly at time scales $t\gtrsim\SI{1}{h}$. (A) Log-log plot of the MSD divided by time of migrating plasmodia distinguished by plasmodium size (7 small, 7 large), and the MSD of the simulated trajectories with strict self-avoidance (large plasmodia) and time-limited self-avoidance of $\SI{3}{h}$ (small plasmodia). Inset images: Example of a large and a small plasmodium, scale bars $\SI{1}{mm]}$. Shaded regions show the standard deviations. (B) Log-lin plot of the instantaneous MSD exponent $\beta$ in experiments and simulations. (C) Log-log plot of the orientational correlation $C$ in experiments and simulations.}
\label{fig:size}
\end{figure*}

\subsection{Model Is Robust against Environmental Changes}
To test the robustness of our model, we let plasmodia migrate on a nutritious substrate (Materials and Methods). Quantifying the MSD and orientational correlation of the obtained 9 trajectories, we do not observe significant differences from the migration behavior on plain agar on the smallest time scales ($\lesssim\SI{10}{min}$), except for a larger generalized diffusion constant $D$. However, on larger time scales, the exponent of the MSD is higher for migration on nutritious agar, resulting in a higher MSD (Fig.~\ref{fig:nutritious}A,B). The same is true for the orientational correlation (Fig.~\ref{fig:nutritious}C).
 Quantification of the run-and-tumble dynamics as described in the previous section reveals that the running and tumbling durations are still distributed as power laws (\textit{SI Appendix}, Figs.~S2B and S5), but that there are very long runs compared to migration on non-nutritious substrate: the average run duration $\langle R\rangle$ increases from $\SI{48.2}{min}$ to $\SI{103.2}{min}$, while the tumble dynamics do not change significantly, with $\langle T\rangle$ increasing from $\SI{17.9}{min}$ to $\SI{20.6}{min}$ (Materials and Methods, \textit{SI Appendix}, Fig.~S2B). This explains both the larger generalized diffusion constant $D$ and the higher MSD exponent $\beta$ on time scales less than $\SI{1}{h}$, according to our analysis of the influence of the average run duration $\langle R\rangle$ on $D$ and $\beta$.
  The distributions of speeds and turning angles are also largely unaffected, but the plasmodia are larger due to growth induced by the nutritious substrate, which explains the higher MSD exponent $\beta$ on large time scales.
We extract all model parameters from the nutritious agar dataset (\textit{SI Appendix}, Fig.~S6) and run additional simulations with the newly extracted parameters, again yielding an excellent match with the data (Fig.~\ref{fig:nutritious}).
This shows that our model is robust and we set out to investigate as a possible evolutionary advantage of \textit{P.~polycephalum} the influence of the plasmodium size on its space exploration behavior.

\subsection{Self-avoidance Critically Controlled by Organism Size}
We want to experimentally test the prediction of our model that the plasmodium size influences the MSD exponent on time scales larger than $\SI{1}{h}$. The size is an intrinsic property of the plasmodia, which we can simply vary by picking plasmodia of different size. We divide the 14 trajectories on plain agar into two equal groups (7 trajectories each), one containing plasmodia with a cell area smaller than the median cell area ($\SI{0.65}{mm^2}$), and the other containing larger plasmodia. Quantifying these groups separately, we find that the orientational correlation is independent of the organism size at small time scales ($\lesssim\SI{1}{h}$) (Fig.~\ref{fig:size}C).
MSD quantification reveals that large plasmodia have a significantly larger MSD at short time scales, meaning that they have a larger generalized diffusion constant $D$ than small plasmodia (Fig.~\ref{fig:size}A). The reason for this is that they are faster since migration speed correlates positively with cell size \cite{kuroda2015allometry}. 
Most notably, the group of small plasmodia shows a strong decrease of the MSD exponent below $\beta=1.5$ at intermediate and long time scales (Fig.~\ref{fig:size}B), which is not visible in the group of large plasmodia. So, small plasmodia have a lower MSD exponent than large plasmodia at time scales larger than $\SI{1}{h}$ (\textit{SI Appendix}, Fig.~S7), as predicted by our model. 
 Quantification of the run-and-tumble dynamics and cell sizes for both size groups independently shows that large plasmodia have twice the width of the small plasmodia, are 29\% faster and spend 31\% more time in the running phase (Materials and Methods, \textit{SI Appendix} Figs.~S9-11). The other migration characteristics are independent of the size.
 We extract all model parameters from the partitioned data sets and perform additional simulations with the newly extracted parameters. Our model captures the MSD and orientational correlations of the large plasmodia (Fig.~\ref{fig:size}), but overestimates those of the small plasmodia for large time scales (\textit{SI Appendix}, Fig.~S8B).
 This means that the model is missing a factor to accurately describe the migration of small plasmodia.
 By examining the experimental record, we find the reason for their stronger decrease in superdiffusivity: small plasmodia occasionally cross their own trajectories (four out of seven small plasmodia cross their own trajectories one to seven times in around 90 hours, see \textit{SI Appendix}, Figs.~S8A, S9A), revisiting previously visited areas, which makes their space exploration less efficient. 
This is also evident from single-cell MSDs. The individual MSDs of small plasmodia show transient regimes of diffusive motion with an exponent of $\beta=1$, which is not the case for large plasmodia (\textit{SI Appendix}, Figs. S9C and S10C).
 From experimental recordings, we observe that self-crossings in small plasmodia occur not before 3 hours after the deposition of the slime trail (\textit{SI Appendix}, Fig.~S8A), suggesting a limited response to the trail. To test this, we run additional simulations in which plasmodia successively lose responsiveness to parts of the trail that are more than 3 hours in the past, allowing them to revisit regions that they have already visited before the last 3 hours. The simulation results give a good match with the experimental data (Fig.~\ref{fig:size}) and suggest an MSD exponent of $\beta=1$ on long time scales (\textit{SI Appendix}, Fig.~S8B). The parameter describing the response time controls how fast the MSD exponent is converging towards a value of $\beta=1$ on intermediate to long time scales (\textit{SI Appendix}, Fig.~S8B). We also ran simulations with a crossing-associated penalty, but this does not capture the MSD exponent well (\textit{SI Appendix}, Fig.~S8C). To directly test whether the response of plasmodia to the trail is size dependent, we experimentally examine the encounters of small and large plasmodia with the trails of small and large conspecifics. We observe that large plasmodia avoid both the trail of small and large plasmodia, in contrast to small plasmodia which avoid neither the trail of small nor large plasmodia (\textit{SI Appendix}, Figs.~S12 and S13). The failure of small plasmodia to respond to slime trails highlights the importance of size in enabling superdiffusive migration. 

\section{Discussion}
We experimentally investigated and quantified the migration behavior of \textit{P.~polycephalum} plasmodia on plain agar. We found that their migration is characterized by superdiffusion and anomalous persistence on long time scales. Our analysis shows that \textit{P.~polycephalum} performs a self-avoiding run-and-tumble movement. Our data-driven model successfully captures the MSD and the orientational correlation and reveals the individual contributions of the migration parameters. Accounting for a loss of the responsiveness to the trail by small plasmodia, it teaches us that the macroscopic unicellular \textit{P.~polycephalum} achieves superdiffusive migration through size-dependent self-avoidance, which affects the MSD exponent and the orientational correlation on intermediate and long time scales. 

The self-avoiding behavior is achieved by a path-marking mechanism in the form of deposited slime, which acts as an avoidance cue. Path marking as a spatial memory is known not only from Hansel and Gretel, but also from a wide range of organisms. Examples include bacteria that secrete exopolysaccharides \cite{zhao2013psl}, mammalian cells that deposit extracellular matrix components \cite{d2021cell}, or ants that leave pheromone trails \cite{jackson2006communication,robinson2005no}. Trail formation is energetically costly, so its benefits should outweigh its costs to be evolutionarily advantageous \cite{kranz2016effective}. Self-avoiding walks are a beneficial strategy in the sense that they increase the space exploration significantly \cite{barbier2022self}. Special cases of self-avoiding walks have been shown to be optimal, for example, by being more efficient than stochastic trajectories in terms of minimizing the search time \cite{moreau2009chance}, or by generating L\'evy walks \cite{sims2014hierarchical}. \textit{P.~polycephalum} may have developed its path-marking strategy to gain an advantage in terms of space exploration.

We observe that the type of migration behavior remains the same in a nutritious environment, underlining the robustness of our model and suggesting that this is the general mechanism by which \textit{P.~polycephalum} explores its environment. However, the migration dynamics adapt to the environment, resulting in longer run durations and therefore a higher MSD exponent on a nutritious substrate. The adaptation of the migration to the nutritional content of the environment is also observed in other cells, like bacteria \cite{barbara2003marine}, dinoflagellates \cite{bartumeus2003helical}, or \textit{D.~discoideum} \cite{varnum1984effects}. \textit{D.~discoideum} is also known to generate gradients in homogeneous environments through the degradation of nutrients, promoting long-range chemotaxis \cite{tweedy2020seeing}. It is possible that a similar mechanism is responsible for the longer run durations we observe for \textit{P.~polycephalum}, but we leave this for future investigation.

Migration analysis on plain agar with respect to cell size shows that small plasmodia have a lower MSD exponent than large plasmodia on long time scales and are therefore less efficient in terms of space exploration. This is due to small plasmodia crossing the slime trail. Since small plasmodia, unlike large plasmodia, fail to respond to slime trails from both small and large specimens, we suggest that there is a size threshold below which the self-avoidance mechanism no longer works well, perhaps due to a reduced ability to process information. This would also explain the previously observed Brownian motion of very small plasmodia on long time scales \cite{rodiek2015migratory}. 

Our study establishes a link between cell size, long-lasting path-marking and space exploration efficiency via superdiffusion. Increasing size has been hypothesized to be the first evolutionary step in the transition from unicellular to multicellular life \cite{bonner1998origins}. Our findings show that, in the case of \textit{P.~polycephalum}, a larger size is beneficial because it allows superdiffusive space exploration. This advantage could have driven \textit{P.~polycephalum} to form larger and larger cells with many nuclei \cite{gerber2022spatial} and thus to evolve into an organism close to multicellular life.

\begin{acknowledgments}
We thank Raphaël Voituriez, Olivier Bénichou, Agnese Codutti and Eberhard Bodenschatz for helpful discussions. This work was funded by the Deutsche Forschungsgemeinschaft (DFG, German Research Foundation) – AL 1429/5-1 (K.A.), the European Research Council (ERC) under the European Union’s Horizon 2020 research and innovation program (grant agreement No. 947630, FlowMem) (K.A.) and by the Federal Ministry of Education and Research (BMBF) and the Free State of Bavaria under the Excellence Strategy of the Federal Government and the Länder for the TUM Innovation Network “Robot Intelligence in the Synthesis of Life” (K.A.).
\end{acknowledgments}

\appendix

\section{Preparation and Imaging of \textit{P.~polycephalum}}
Plasmodial specimen were prepared from microplasmodia grown in a liquid culture using the medium by Daniel and Rusch \cite{daniel1961pure} with hematin instead of chicken embryo extract \cite{daniel1962hematin}. Culture medium containing microplasmodia was pipetted onto an agar plate, from which individual microplasmodia with a diameter between 0.5 and $\SI{1}{mm}$ were selected with a $\SI{1}{mL}$ pipette tip and transferred to another 1.5\% agar plate, with a resulting plasmodial area density of <0.01. Nutritious agar contained 10\%  of the culture medium. 

Plasmodia (0.2\,-\,$\SI{2}{\mathrm{mm}^2}$ cell area) were imaged directly after plating under microscope light for 72 to 168 hours, with controlled temperature (24.4\degree C) and humidity (close to saturation). The plasmodia were imaged with a Zeiss Axio Zoom V.16 microscope equipped with a custom-made Pecon stage-top incubation system, a Hamamatsu ORCA-Flash 4.0 digital camera and a Zeiss PlanApo Z 0.5x objective, yielding a resolution of $\SI{10.83}{\micro\meter}$/pixel. A green filter ($550/\SI{50}{nm}$) was placed over the transmission light source of the microscope to diminish \textit{P.~polycephalum}`s response to the continuous illumination with light, since these wavelengths are known not to induce phototaxis in \textit{P.~polycephalum} \cite{hato1976phototaxis}. In this way, plasmodia were exposed to a light intensity of only $\SI{0.078}{W\per\square\mm}$. Zeiss Zen 3.2 (blue edition) software was used for imaging. A tiled image composed of 16 tiles was acquired every $\SI{4}{min}$ using the Zen Tiles tool, yielding a field of view of 8.1\,x\,$\SI{8.1}{cm}$. A total number of eight experiments was conducted, for examples see Supplementary Movies S1 and S2.

\section{Image processing and analysis}
All tiles were converted into 8-bit TIFF files using Zeiss Zen 3.2 and their backgrounds were removed with a rolling ball algorithm. The MIST stitching algorithm \cite{blattner2014hybrid,chalfoun2017mist} was used to assemble the tiles into a single image. From the stitched images, we extracted plasmodium pixels using a custom-written MATLAB (The MathWorks) code, generating binary images via intensity thresholding. Subsequently, we calculated the center of mass of each plasmodium in each image to obtain the trajectories. For sufficient statistics, only trajectories with a length of at least $\SI{30}{h}$ were considered for the analysis. 

All mean squared displacements in the main text are ensemble averages of the time-averaged squared displacements of individual trajectories. The same approach was used to calculate the orientational correlations. The instantaneous MSD exponent $\beta$ is the logarithmic derivative of the MSD. Its standard deviation is computed via error propagation (\textit{SI Appendix}, supporting text). Data is shown up to the point where the standard deviation of $\beta$ exceeds the boundaries given by $\beta=1$ and $\beta=2$. 

Tumbling events were identified by calculating the directionality \cite{varennes2019physical} of the migration as the ratio of the plasmodium’s displacement to the total distance traveled within a time frame of 20 minutes. If the directionality was higher than the threshold value 0.9, the time point was attributed to a running event, else to a tumbling event (\textit{SI Appendix}, Fig.~S1). The model parameters are robust against deviations of 5\% from this threshold (\textit{SI Appendix}, Figs.~S14). Note that, in principle, tumbles could be caused by encountering a slime trail, which would affect the distribution of run durations. However, we observe that only 2.4\% of runs longer than $\SI{28}{\min}$ are stopped by an encounter with a slime trail -- a fraction too small to affect the distributions significantly.

The typical width of a plasmodium was estimated by the median length of the minor axis of the ellipse that has the same second moment of area as the plasmodium, which was calculated via the built-in MATLAB function \textit{regionprops} using the property \textit{MinorAxisLength}.

\section{Simulations}
We simulated trajectories as a series of runs and tumbles, which themselves consist of series of random steps characterized by a direction and by a step length. The number of steps during a run or a tumble is determined by drawing a random number from a combination of two power laws (\textit{SI Appendix, supporting text}) with parameters $\lambda_\mathrm{R,T}$ and $k_\mathrm{R,T}$.
During tumbles, directions were determined by drawing a random angle from a uniform distribution between [$-\pi$,$\pi$], and speeds were fixed to $v_\mathrm{T}$. During runs, directions were determined by drawing a random angle from an exponential distribution with mean $\phi_\mathrm{R}$, and speeds were fixed to $v_\mathrm{R}$, with the additional rule that a step can only be made if it does not come closer to the previous trajectory than twice the mean plasmodium width $\bar{w}$. This rule enforces self-avoidance. All parameters used in the simulations were estimated directly from the experimental data by fitting the distributions described in the main text to the experimental data (see Table \ref{table:values}).

Only simulated trajectories with a minimum length of 1800 steps, corresponding to a migration duration of $\SI{120}{h}$, were selected to ensure the true asymptotic behavior (\textit{SI Appendix, supporting text} and Fig.~S15). All statistics were averaged over 5000 simulated trajectories each. Simulations were performed using Python.

\setlength{\tabcolsep}{1.4pt}
\begin{table}[h!]
\caption{Model parameter values inferred from  data sets.}
\begin{ruledtabular}
\begin{tabular}{l|c|c|c|c|c|c}
 Data & $\alpha_\mathrm{R,T}$ & $\chi_\mathrm{R,T}$ & $b_\mathrm{R,T}$ [min] & $v_\mathrm{R,T} \left[\frac{\si{\micro\metre}}{\mathrm{min}}\right]$ & $\phi_\mathrm{R}$  [rad] & $\bar{w}$ [µm]\\
\hline
PA  & 1.1, 0.2  & 2.8, 5.3  & 98, 28    & 24.5, 12.9    & 0.24 & 954\\
NA  & 1.1, 0.8  & 2.2, 4.1  & 132, 36   & 24.9, 16.7    & 0.25 & 1811\\
SP  & 1.2, 0.03 & 2.9, 3.6  & 96, 20    & 21.6, 11.8    & 0.25 & 674\\
LP  & 1.0, 0.2  & 3.1, 4.6  & 124, 28   & 27.9, 14.3    & 0.23 & 1340\\
\end{tabular}
\end{ruledtabular}
\blfootnote{PA/NA = plain/nutritious agar, SP/LP = small/large plasmodia.}
\label{table:values}
\end{table}%

\section{Data Availability}
Microscopy recordings data have been deposited in mediaTUM (\url{https://mediatum.ub.tum.de/1734713}).

\bibliography{bibliography}

\begin{thebibliography}{59}%
\makeatletter
\providecommand \@ifxundefined [1]{%
 \@ifx{#1\undefined}
}%
\providecommand \@ifnum [1]{%
 \ifnum #1\expandafter \@firstoftwo
 \else \expandafter \@secondoftwo
 \fi
}%
\providecommand \@ifx [1]{%
 \ifx #1\expandafter \@firstoftwo
 \else \expandafter \@secondoftwo
 \fi
}%
\providecommand \natexlab [1]{#1}%
\providecommand \enquote  [1]{``#1''}%
\providecommand \bibnamefont  [1]{#1}%
\providecommand \bibfnamefont [1]{#1}%
\providecommand \citenamefont [1]{#1}%
\providecommand \href@noop [0]{\@secondoftwo}%
\providecommand \href [0]{\begingroup \@sanitize@url \@href}%
\providecommand \@href[1]{\@@startlink{#1}\@@href}%
\providecommand \@@href[1]{\endgroup#1\@@endlink}%
\providecommand \@sanitize@url [0]{\catcode `\\12\catcode `\$12\catcode
  `\&12\catcode `\#12\catcode `\^12\catcode `\_12\catcode `\%12\relax}%
\providecommand \@@startlink[1]{}%
\providecommand \@@endlink[0]{}%
\providecommand \url  [0]{\begingroup\@sanitize@url \@url }%
\providecommand \@url [1]{\endgroup\@href {#1}{\urlprefix }}%
\providecommand \urlprefix  [0]{URL }%
\providecommand \Eprint [0]{\href }%
\providecommand \doibase [0]{https://doi.org/}%
\providecommand \selectlanguage [0]{\@gobble}%
\providecommand \bibinfo  [0]{\@secondoftwo}%
\providecommand \bibfield  [0]{\@secondoftwo}%
\providecommand \translation [1]{[#1]}%
\providecommand \BibitemOpen [0]{}%
\providecommand \bibitemStop [0]{}%
\providecommand \bibitemNoStop [0]{.\EOS\space}%
\providecommand \EOS [0]{\spacefactor3000\relax}%
\providecommand \BibitemShut  [1]{\csname bibitem#1\endcsname}%
\let\auto@bib@innerbib\@empty
\bibitem [{\citenamefont {Berg}\ and\ \citenamefont
  {Brown}(1972)}]{berg1972chemotaxis}%
  \BibitemOpen
  \bibfield  {author} {\bibinfo {author} {\bibfnamefont {H.~C.}\ \bibnamefont
  {Berg}}\ and\ \bibinfo {author} {\bibfnamefont {D.~A.}\ \bibnamefont
  {Brown}},\ }\bibfield  {title} {\bibinfo {title} {Chemotaxis in
  \textit{Escherichia coli} analysed by three-dimensional tracking},\
  }\href@noop {} {\bibfield  {journal} {\bibinfo  {journal} {Nature}\ }\textbf
  {\bibinfo {volume} {239}},\ \bibinfo {pages} {500} (\bibinfo {year}
  {1972})}\BibitemShut {NoStop}%
\bibitem [{\citenamefont {Li}\ \emph {et~al.}(2008)\citenamefont {Li},
  \citenamefont {N{\o}rrelykke},\ and\ \citenamefont {Cox}}]{li2008persistent}%
  \BibitemOpen
  \bibfield  {author} {\bibinfo {author} {\bibfnamefont {L.}~\bibnamefont
  {Li}}, \bibinfo {author} {\bibfnamefont {S.~F.}\ \bibnamefont
  {N{\o}rrelykke}},\ and\ \bibinfo {author} {\bibfnamefont {E.~C.}\
  \bibnamefont {Cox}},\ }\bibfield  {title} {\bibinfo {title} {Persistent cell
  motion in the absence of external signals: A search strategy for eukaryotic
  cells},\ }\href@noop {} {\bibfield  {journal} {\bibinfo  {journal} {PLoS
  one}\ }\textbf {\bibinfo {volume} {3}},\ \bibinfo {pages} {e2093} (\bibinfo
  {year} {2008})}\BibitemShut {NoStop}%
\bibitem [{\citenamefont {Jackson}\ and\ \citenamefont
  {Ratnieks}(2006)}]{jackson2006communication}%
  \BibitemOpen
  \bibfield  {author} {\bibinfo {author} {\bibfnamefont {D.~E.}\ \bibnamefont
  {Jackson}}\ and\ \bibinfo {author} {\bibfnamefont {F.~L.}\ \bibnamefont
  {Ratnieks}},\ }\bibfield  {title} {\bibinfo {title} {Communication in ants},\
  }\href@noop {} {\bibfield  {journal} {\bibinfo  {journal} {Current biology}\
  }\textbf {\bibinfo {volume} {16}},\ \bibinfo {pages} {R570} (\bibinfo {year}
  {2006})}\BibitemShut {NoStop}%
\bibitem [{\citenamefont {Sims}\ \emph {et~al.}(2008)\citenamefont {Sims},
  \citenamefont {Southall}, \citenamefont {Humphries}, \citenamefont {Hays},
  \citenamefont {Bradshaw}, \citenamefont {Pitchford}, \citenamefont {James},
  \citenamefont {Ahmed}, \citenamefont {Brierley}, \citenamefont {Hindell}
  \emph {et~al.}}]{sims2008scaling}%
  \BibitemOpen
  \bibfield  {author} {\bibinfo {author} {\bibfnamefont {D.~W.}\ \bibnamefont
  {Sims}}, \bibinfo {author} {\bibfnamefont {E.~J.}\ \bibnamefont {Southall}},
  \bibinfo {author} {\bibfnamefont {N.~E.}\ \bibnamefont {Humphries}}, \bibinfo
  {author} {\bibfnamefont {G.~C.}\ \bibnamefont {Hays}}, \bibinfo {author}
  {\bibfnamefont {C.~J.}\ \bibnamefont {Bradshaw}}, \bibinfo {author}
  {\bibfnamefont {J.~W.}\ \bibnamefont {Pitchford}}, \bibinfo {author}
  {\bibfnamefont {A.}~\bibnamefont {James}}, \bibinfo {author} {\bibfnamefont
  {M.~Z.}\ \bibnamefont {Ahmed}}, \bibinfo {author} {\bibfnamefont {A.~S.}\
  \bibnamefont {Brierley}}, \bibinfo {author} {\bibfnamefont {M.~A.}\
  \bibnamefont {Hindell}}, \emph {et~al.},\ }\bibfield  {title} {\bibinfo
  {title} {Scaling laws of marine predator search behaviour},\ }\href@noop {}
  {\bibfield  {journal} {\bibinfo  {journal} {Nature}\ }\textbf {\bibinfo
  {volume} {451}},\ \bibinfo {pages} {1098} (\bibinfo {year}
  {2008})}\BibitemShut {NoStop}%
\bibitem [{\citenamefont {Raichlen}\ \emph {et~al.}(2014)\citenamefont
  {Raichlen}, \citenamefont {Wood}, \citenamefont {Gordon}, \citenamefont
  {Mabulla}, \citenamefont {Marlowe},\ and\ \citenamefont
  {Pontzer}}]{raichlen2014evidence}%
  \BibitemOpen
  \bibfield  {author} {\bibinfo {author} {\bibfnamefont {D.~A.}\ \bibnamefont
  {Raichlen}}, \bibinfo {author} {\bibfnamefont {B.~M.}\ \bibnamefont {Wood}},
  \bibinfo {author} {\bibfnamefont {A.~D.}\ \bibnamefont {Gordon}}, \bibinfo
  {author} {\bibfnamefont {A.~Z.}\ \bibnamefont {Mabulla}}, \bibinfo {author}
  {\bibfnamefont {F.~W.}\ \bibnamefont {Marlowe}},\ and\ \bibinfo {author}
  {\bibfnamefont {H.}~\bibnamefont {Pontzer}},\ }\bibfield  {title} {\bibinfo
  {title} {Evidence of \text{L{\'e}vy} walk foraging patterns in human
  hunter--gatherers},\ }\href@noop {} {\bibfield  {journal} {\bibinfo
  {journal} {Proceedings of the National Academy of Sciences}\ }\textbf
  {\bibinfo {volume} {111}},\ \bibinfo {pages} {728} (\bibinfo {year}
  {2014})}\BibitemShut {NoStop}%
\bibitem [{\citenamefont {Bartumeus}\ \emph {et~al.}(2005)\citenamefont
  {Bartumeus}, \citenamefont {da~Luz}, \citenamefont {Viswanathan},\ and\
  \citenamefont {Catalan}}]{bartumeus2005animal}%
  \BibitemOpen
  \bibfield  {author} {\bibinfo {author} {\bibfnamefont {F.}~\bibnamefont
  {Bartumeus}}, \bibinfo {author} {\bibfnamefont {M.~G.~E.}\ \bibnamefont
  {da~Luz}}, \bibinfo {author} {\bibfnamefont {G.~M.}\ \bibnamefont
  {Viswanathan}},\ and\ \bibinfo {author} {\bibfnamefont {J.}~\bibnamefont
  {Catalan}},\ }\bibfield  {title} {\bibinfo {title} {Animal search strategies:
  a quantitative random-walk analysis},\ }\href@noop {} {\bibfield  {journal}
  {\bibinfo  {journal} {Ecology}\ }\textbf {\bibinfo {volume} {86}},\ \bibinfo
  {pages} {3078} (\bibinfo {year} {2005})}\BibitemShut {NoStop}%
\bibitem [{\citenamefont {Shaebani}\ \emph {et~al.}(2022)\citenamefont
  {Shaebani}, \citenamefont {Piel},\ and\ \citenamefont
  {Lautenschl{\"a}ger}}]{shaebani2022distinct}%
  \BibitemOpen
  \bibfield  {author} {\bibinfo {author} {\bibfnamefont {M.~R.}\ \bibnamefont
  {Shaebani}}, \bibinfo {author} {\bibfnamefont {M.}~\bibnamefont {Piel}},\
  and\ \bibinfo {author} {\bibfnamefont {F.}~\bibnamefont
  {Lautenschl{\"a}ger}},\ }\bibfield  {title} {\bibinfo {title} {Distinct speed
  and direction memories of migrating dendritic cells diversify their search
  strategies},\ }\href@noop {} {\bibfield  {journal} {\bibinfo  {journal}
  {Biophysical Journal}\ }\textbf {\bibinfo {volume} {121}},\ \bibinfo {pages}
  {4099} (\bibinfo {year} {2022})}\BibitemShut {NoStop}%
\bibitem [{\citenamefont {Palyulin}\ \emph {et~al.}(2014)\citenamefont
  {Palyulin}, \citenamefont {Chechkin},\ and\ \citenamefont
  {Metzler}}]{palyulin2014levy}%
  \BibitemOpen
  \bibfield  {author} {\bibinfo {author} {\bibfnamefont {V.~V.}\ \bibnamefont
  {Palyulin}}, \bibinfo {author} {\bibfnamefont {A.~V.}\ \bibnamefont
  {Chechkin}},\ and\ \bibinfo {author} {\bibfnamefont {R.}~\bibnamefont
  {Metzler}},\ }\bibfield  {title} {\bibinfo {title} {L{\'e}vy flights do not
  always optimize random blind search for sparse targets},\ }\href@noop {}
  {\bibfield  {journal} {\bibinfo  {journal} {Proceedings of the National
  Academy of Sciences}\ }\textbf {\bibinfo {volume} {111}},\ \bibinfo {pages}
  {2931} (\bibinfo {year} {2014})}\BibitemShut {NoStop}%
\bibitem [{\citenamefont {Faustino}\ \emph {et~al.}(2007)\citenamefont
  {Faustino}, \citenamefont {Da~Silva}, \citenamefont {Da~Luz}, \citenamefont
  {Raposo},\ and\ \citenamefont {Viswanathan}}]{faustino2007search}%
  \BibitemOpen
  \bibfield  {author} {\bibinfo {author} {\bibfnamefont {C.}~\bibnamefont
  {Faustino}}, \bibinfo {author} {\bibfnamefont {L.}~\bibnamefont {Da~Silva}},
  \bibinfo {author} {\bibfnamefont {M.}~\bibnamefont {Da~Luz}}, \bibinfo
  {author} {\bibfnamefont {E.}~\bibnamefont {Raposo}},\ and\ \bibinfo {author}
  {\bibfnamefont {G.}~\bibnamefont {Viswanathan}},\ }\bibfield  {title}
  {\bibinfo {title} {Search dynamics at the edge of extinction: Anomalous
  diffusion as a critical survival state},\ }\href@noop {} {\bibfield
  {journal} {\bibinfo  {journal} {EPL (Europhysics Letters)}\ }\textbf
  {\bibinfo {volume} {77}},\ \bibinfo {pages} {30002} (\bibinfo {year}
  {2007})}\BibitemShut {NoStop}%
\bibitem [{\citenamefont {Rupprecht}\ \emph {et~al.}(2016)\citenamefont
  {Rupprecht}, \citenamefont {B{\'e}nichou},\ and\ \citenamefont
  {Voituriez}}]{rupprecht2016optimal}%
  \BibitemOpen
  \bibfield  {author} {\bibinfo {author} {\bibfnamefont {J.-F.}\ \bibnamefont
  {Rupprecht}}, \bibinfo {author} {\bibfnamefont {O.}~\bibnamefont
  {B{\'e}nichou}},\ and\ \bibinfo {author} {\bibfnamefont {R.}~\bibnamefont
  {Voituriez}},\ }\bibfield  {title} {\bibinfo {title} {Optimal search
  strategies of run-and-tumble walks},\ }\href@noop {} {\bibfield  {journal}
  {\bibinfo  {journal} {Physical Review E}\ }\textbf {\bibinfo {volume} {94}},\
  \bibinfo {pages} {012117} (\bibinfo {year} {2016})}\BibitemShut {NoStop}%
\bibitem [{\citenamefont {F{\"u}rth}(1920)}]{furth1920brownsche}%
  \BibitemOpen
  \bibfield  {author} {\bibinfo {author} {\bibfnamefont {R.}~\bibnamefont
  {F{\"u}rth}},\ }\bibfield  {title} {\bibinfo {title} {Die \text{Brownsche}
  \text{Bewegung} bei \text{Ber{\"u}cksichtigung} einer \text{Persistenz} der
  \text{Bewegungsrichtung}. \text{Mit} \text{Anwendungen} auf die
  \text{Bewegung} lebender \text{Infusorien}},\ }\href@noop {} {\bibfield
  {journal} {\bibinfo  {journal} {Zeitschrift f{\"u}r Physik}\ }\textbf
  {\bibinfo {volume} {2}},\ \bibinfo {pages} {244} (\bibinfo {year}
  {1920})}\BibitemShut {NoStop}%
\bibitem [{\citenamefont {Wu}\ \emph {et~al.}(2014)\citenamefont {Wu},
  \citenamefont {Giri}, \citenamefont {Sun},\ and\ \citenamefont
  {Wirtz}}]{wu2014three}%
  \BibitemOpen
  \bibfield  {author} {\bibinfo {author} {\bibfnamefont {P.-H.}\ \bibnamefont
  {Wu}}, \bibinfo {author} {\bibfnamefont {A.}~\bibnamefont {Giri}}, \bibinfo
  {author} {\bibfnamefont {S.~X.}\ \bibnamefont {Sun}},\ and\ \bibinfo {author}
  {\bibfnamefont {D.}~\bibnamefont {Wirtz}},\ }\bibfield  {title} {\bibinfo
  {title} {Three-dimensional cell migration does not follow a random walk},\
  }\href@noop {} {\bibfield  {journal} {\bibinfo  {journal} {Proceedings of the
  National Academy of Sciences}\ }\textbf {\bibinfo {volume} {111}},\ \bibinfo
  {pages} {3949} (\bibinfo {year} {2014})}\BibitemShut {NoStop}%
\bibitem [{\citenamefont {Dieterich}\ \emph {et~al.}(2008)\citenamefont
  {Dieterich}, \citenamefont {Klages}, \citenamefont {Preuss},\ and\
  \citenamefont {Schwab}}]{dieterich2008anomalous}%
  \BibitemOpen
  \bibfield  {author} {\bibinfo {author} {\bibfnamefont {P.}~\bibnamefont
  {Dieterich}}, \bibinfo {author} {\bibfnamefont {R.}~\bibnamefont {Klages}},
  \bibinfo {author} {\bibfnamefont {R.}~\bibnamefont {Preuss}},\ and\ \bibinfo
  {author} {\bibfnamefont {A.}~\bibnamefont {Schwab}},\ }\bibfield  {title}
  {\bibinfo {title} {Anomalous dynamics of cell migration},\ }\href@noop {}
  {\bibfield  {journal} {\bibinfo  {journal} {Proceedings of the National
  Academy of Sciences}\ }\textbf {\bibinfo {volume} {105}},\ \bibinfo {pages}
  {459} (\bibinfo {year} {2008})}\BibitemShut {NoStop}%
\bibitem [{\citenamefont {Harris}\ \emph {et~al.}(2012)\citenamefont {Harris},
  \citenamefont {Banigan}, \citenamefont {Christian}, \citenamefont {Konradt},
  \citenamefont {Tait~Wojno}, \citenamefont {Norose}, \citenamefont {Wilson},
  \citenamefont {John}, \citenamefont {Weninger}, \citenamefont {Luster} \emph
  {et~al.}}]{harris2012generalized}%
  \BibitemOpen
  \bibfield  {author} {\bibinfo {author} {\bibfnamefont {T.~H.}\ \bibnamefont
  {Harris}}, \bibinfo {author} {\bibfnamefont {E.~J.}\ \bibnamefont {Banigan}},
  \bibinfo {author} {\bibfnamefont {D.~A.}\ \bibnamefont {Christian}}, \bibinfo
  {author} {\bibfnamefont {C.}~\bibnamefont {Konradt}}, \bibinfo {author}
  {\bibfnamefont {E.~D.}\ \bibnamefont {Tait~Wojno}}, \bibinfo {author}
  {\bibfnamefont {K.}~\bibnamefont {Norose}}, \bibinfo {author} {\bibfnamefont
  {E.~H.}\ \bibnamefont {Wilson}}, \bibinfo {author} {\bibfnamefont
  {B.}~\bibnamefont {John}}, \bibinfo {author} {\bibfnamefont {W.}~\bibnamefont
  {Weninger}}, \bibinfo {author} {\bibfnamefont {A.~D.}\ \bibnamefont
  {Luster}}, \emph {et~al.},\ }\bibfield  {title} {\bibinfo {title}
  {Generalized \text{L{\'e}vy} walks and the role of chemokines in migration of
  effector cd8+ t cells},\ }\href@noop {} {\bibfield  {journal} {\bibinfo
  {journal} {Nature}\ }\textbf {\bibinfo {volume} {486}},\ \bibinfo {pages}
  {545} (\bibinfo {year} {2012})}\BibitemShut {NoStop}%
\bibitem [{\citenamefont {Huda}\ \emph {et~al.}(2018)\citenamefont {Huda},
  \citenamefont {Weigelin}, \citenamefont {Wolf}, \citenamefont {Tretiakov},
  \citenamefont {Polev}, \citenamefont {Wilk}, \citenamefont {Iwasa},
  \citenamefont {Emami}, \citenamefont {Narojczyk}, \citenamefont {Banaszak}
  \emph {et~al.}}]{huda2018levy}%
  \BibitemOpen
  \bibfield  {author} {\bibinfo {author} {\bibfnamefont {S.}~\bibnamefont
  {Huda}}, \bibinfo {author} {\bibfnamefont {B.}~\bibnamefont {Weigelin}},
  \bibinfo {author} {\bibfnamefont {K.}~\bibnamefont {Wolf}}, \bibinfo {author}
  {\bibfnamefont {K.~V.}\ \bibnamefont {Tretiakov}}, \bibinfo {author}
  {\bibfnamefont {K.}~\bibnamefont {Polev}}, \bibinfo {author} {\bibfnamefont
  {G.}~\bibnamefont {Wilk}}, \bibinfo {author} {\bibfnamefont {M.}~\bibnamefont
  {Iwasa}}, \bibinfo {author} {\bibfnamefont {F.~S.}\ \bibnamefont {Emami}},
  \bibinfo {author} {\bibfnamefont {J.~W.}\ \bibnamefont {Narojczyk}}, \bibinfo
  {author} {\bibfnamefont {M.}~\bibnamefont {Banaszak}}, \emph {et~al.},\
  }\bibfield  {title} {\bibinfo {title} {L{\'e}vy-like movement patterns of
  metastatic cancer cells revealed in microfabricated systems and implicated in
  vivo},\ }\href@noop {} {\bibfield  {journal} {\bibinfo  {journal} {Nature
  communications}\ }\textbf {\bibinfo {volume} {9}},\ \bibinfo {pages} {1}
  (\bibinfo {year} {2018})}\BibitemShut {NoStop}%
\bibitem [{\citenamefont {Upadhyaya}\ \emph {et~al.}(2001)\citenamefont
  {Upadhyaya}, \citenamefont {Rieu}, \citenamefont {Glazier},\ and\
  \citenamefont {Sawada}}]{upadhyaya2001anomalous}%
  \BibitemOpen
  \bibfield  {author} {\bibinfo {author} {\bibfnamefont {A.}~\bibnamefont
  {Upadhyaya}}, \bibinfo {author} {\bibfnamefont {J.-P.}\ \bibnamefont {Rieu}},
  \bibinfo {author} {\bibfnamefont {J.~A.}\ \bibnamefont {Glazier}},\ and\
  \bibinfo {author} {\bibfnamefont {Y.}~\bibnamefont {Sawada}},\ }\bibfield
  {title} {\bibinfo {title} {Anomalous diffusion and non-gaussian velocity
  distribution of \textit{Hydra} cells in cellular aggregates},\ }\href@noop {}
  {\bibfield  {journal} {\bibinfo  {journal} {Physica A: Statistical Mechanics
  and its Applications}\ }\textbf {\bibinfo {volume} {293}},\ \bibinfo {pages}
  {549} (\bibinfo {year} {2001})}\BibitemShut {NoStop}%
\bibitem [{\citenamefont {Miramontes}\ \emph {et~al.}(2014)\citenamefont
  {Miramontes}, \citenamefont {DeSouza}, \citenamefont {Paiva}, \citenamefont
  {Marins},\ and\ \citenamefont {Orozco}}]{miramontes2014levy}%
  \BibitemOpen
  \bibfield  {author} {\bibinfo {author} {\bibfnamefont {O.}~\bibnamefont
  {Miramontes}}, \bibinfo {author} {\bibfnamefont {O.}~\bibnamefont {DeSouza}},
  \bibinfo {author} {\bibfnamefont {L.~R.}\ \bibnamefont {Paiva}}, \bibinfo
  {author} {\bibfnamefont {A.}~\bibnamefont {Marins}},\ and\ \bibinfo {author}
  {\bibfnamefont {S.}~\bibnamefont {Orozco}},\ }\bibfield  {title} {\bibinfo
  {title} {L{\'e}vy flights and self-similar exploratory behaviour of termite
  workers: Beyond model fitting},\ }\href@noop {} {\bibfield  {journal}
  {\bibinfo  {journal} {PloS one}\ }\textbf {\bibinfo {volume} {9}},\ \bibinfo
  {pages} {e111183} (\bibinfo {year} {2014})}\BibitemShut {NoStop}%
\bibitem [{\citenamefont {Viswanathan}\ \emph {et~al.}(1996)\citenamefont
  {Viswanathan}, \citenamefont {Afanasyev}, \citenamefont {Buldyrev},
  \citenamefont {Murphy}, \citenamefont {Prince},\ and\ \citenamefont
  {Stanley}}]{viswanathan1996levy}%
  \BibitemOpen
  \bibfield  {author} {\bibinfo {author} {\bibfnamefont {G.~M.}\ \bibnamefont
  {Viswanathan}}, \bibinfo {author} {\bibfnamefont {V.}~\bibnamefont
  {Afanasyev}}, \bibinfo {author} {\bibfnamefont {S.~V.}\ \bibnamefont
  {Buldyrev}}, \bibinfo {author} {\bibfnamefont {E.~J.}\ \bibnamefont
  {Murphy}}, \bibinfo {author} {\bibfnamefont {P.~A.}\ \bibnamefont {Prince}},\
  and\ \bibinfo {author} {\bibfnamefont {H.~E.}\ \bibnamefont {Stanley}},\
  }\bibfield  {title} {\bibinfo {title} {L{\'e}vy flight search patterns of
  wandering albatrosses},\ }\href@noop {} {\bibfield  {journal} {\bibinfo
  {journal} {Nature}\ }\textbf {\bibinfo {volume} {381}},\ \bibinfo {pages}
  {413} (\bibinfo {year} {1996})}\BibitemShut {NoStop}%
\bibitem [{\citenamefont {Vilk}\ \emph {et~al.}(2022)\citenamefont {Vilk},
  \citenamefont {Orchan}, \citenamefont {Charter}, \citenamefont {Ganot},
  \citenamefont {Toledo}, \citenamefont {Nathan},\ and\ \citenamefont
  {Assaf}}]{vilk2022ergodicity}%
  \BibitemOpen
  \bibfield  {author} {\bibinfo {author} {\bibfnamefont {O.}~\bibnamefont
  {Vilk}}, \bibinfo {author} {\bibfnamefont {Y.}~\bibnamefont {Orchan}},
  \bibinfo {author} {\bibfnamefont {M.}~\bibnamefont {Charter}}, \bibinfo
  {author} {\bibfnamefont {N.}~\bibnamefont {Ganot}}, \bibinfo {author}
  {\bibfnamefont {S.}~\bibnamefont {Toledo}}, \bibinfo {author} {\bibfnamefont
  {R.}~\bibnamefont {Nathan}},\ and\ \bibinfo {author} {\bibfnamefont
  {M.}~\bibnamefont {Assaf}},\ }\bibfield  {title} {\bibinfo {title}
  {Ergodicity breaking in area-restricted search of avian predators},\
  }\href@noop {} {\bibfield  {journal} {\bibinfo  {journal} {Physical Review
  X}\ }\textbf {\bibinfo {volume} {12}},\ \bibinfo {pages} {031005} (\bibinfo
  {year} {2022})}\BibitemShut {NoStop}%
\bibitem [{\citenamefont {Ariel}\ \emph {et~al.}(2015)\citenamefont {Ariel},
  \citenamefont {Rabani}, \citenamefont {Benisty}, \citenamefont {Partridge},
  \citenamefont {Harshey},\ and\ \citenamefont {Be'Er}}]{ariel2015swarming}%
  \BibitemOpen
  \bibfield  {author} {\bibinfo {author} {\bibfnamefont {G.}~\bibnamefont
  {Ariel}}, \bibinfo {author} {\bibfnamefont {A.}~\bibnamefont {Rabani}},
  \bibinfo {author} {\bibfnamefont {S.}~\bibnamefont {Benisty}}, \bibinfo
  {author} {\bibfnamefont {J.~D.}\ \bibnamefont {Partridge}}, \bibinfo {author}
  {\bibfnamefont {R.~M.}\ \bibnamefont {Harshey}},\ and\ \bibinfo {author}
  {\bibfnamefont {A.}~\bibnamefont {Be'Er}},\ }\bibfield  {title} {\bibinfo
  {title} {Swarming bacteria migrate by \text{L{\'e}vy} walk},\ }\href@noop {}
  {\bibfield  {journal} {\bibinfo  {journal} {Nature communications}\ }\textbf
  {\bibinfo {volume} {6}},\ \bibinfo {pages} {8396} (\bibinfo {year}
  {2015})}\BibitemShut {NoStop}%
\bibitem [{\citenamefont {Rodiek}\ and\ \citenamefont
  {Hauser}(2015)}]{rodiek2015migratory}%
  \BibitemOpen
  \bibfield  {author} {\bibinfo {author} {\bibfnamefont {B.}~\bibnamefont
  {Rodiek}}\ and\ \bibinfo {author} {\bibfnamefont {M.}~\bibnamefont
  {Hauser}},\ }\bibfield  {title} {\bibinfo {title} {Migratory behaviour of
  \textit{Physarum polycephalum} microplasmodia},\ }\href@noop {} {\bibfield
  {journal} {\bibinfo  {journal} {The European Physical Journal Special
  Topics}\ }\textbf {\bibinfo {volume} {224}},\ \bibinfo {pages} {1199}
  (\bibinfo {year} {2015})}\BibitemShut {NoStop}%
\bibitem [{\citenamefont {Shirakawa}\ \emph {et~al.}(2019)\citenamefont
  {Shirakawa}, \citenamefont {Niizato}, \citenamefont {Sato},\ and\
  \citenamefont {Ohno}}]{shirakawa2019biased}%
  \BibitemOpen
  \bibfield  {author} {\bibinfo {author} {\bibfnamefont {T.}~\bibnamefont
  {Shirakawa}}, \bibinfo {author} {\bibfnamefont {T.}~\bibnamefont {Niizato}},
  \bibinfo {author} {\bibfnamefont {H.}~\bibnamefont {Sato}},\ and\ \bibinfo
  {author} {\bibfnamefont {R.}~\bibnamefont {Ohno}},\ }\bibfield  {title}
  {\bibinfo {title} {Biased \text{L{\'e}vy}-walk pattern in the exploratory
  behavior of the \textit{Physarum} plasmodium},\ }\href@noop {} {\bibfield
  {journal} {\bibinfo  {journal} {Biosystems}\ }\textbf {\bibinfo {volume}
  {182}},\ \bibinfo {pages} {52} (\bibinfo {year} {2019})}\BibitemShut
  {NoStop}%
\bibitem [{\citenamefont {Bonner}(1998)}]{bonner1998origins}%
  \BibitemOpen
  \bibfield  {author} {\bibinfo {author} {\bibfnamefont {J.~T.}\ \bibnamefont
  {Bonner}},\ }\bibfield  {title} {\bibinfo {title} {The origins of
  multicellularity},\ }\href@noop {} {\bibfield  {journal} {\bibinfo  {journal}
  {Integrative Biology: Issues, News, and Reviews: Published in Association
  with The Society for Integrative and Comparative Biology}\ }\textbf {\bibinfo
  {volume} {1}},\ \bibinfo {pages} {27} (\bibinfo {year} {1998})}\BibitemShut
  {NoStop}%
\bibitem [{\citenamefont {Metzler}\ \emph {et~al.}(2014)\citenamefont
  {Metzler}, \citenamefont {Jeon}, \citenamefont {Cherstvy},\ and\
  \citenamefont {Barkai}}]{metzler2014anomalous}%
  \BibitemOpen
  \bibfield  {author} {\bibinfo {author} {\bibfnamefont {R.}~\bibnamefont
  {Metzler}}, \bibinfo {author} {\bibfnamefont {J.-H.}\ \bibnamefont {Jeon}},
  \bibinfo {author} {\bibfnamefont {A.~G.}\ \bibnamefont {Cherstvy}},\ and\
  \bibinfo {author} {\bibfnamefont {E.}~\bibnamefont {Barkai}},\ }\bibfield
  {title} {\bibinfo {title} {Anomalous diffusion models and their properties:
  non-stationarity, non-ergodicity, and ageing at the centenary of single
  particle tracking},\ }\href@noop {} {\bibfield  {journal} {\bibinfo
  {journal} {Physical Chemistry Chemical Physics}\ }\textbf {\bibinfo {volume}
  {16}},\ \bibinfo {pages} {24128} (\bibinfo {year} {2014})}\BibitemShut
  {NoStop}%
\bibitem [{\citenamefont {Viswanathan}\ \emph {et~al.}(2008)\citenamefont
  {Viswanathan}, \citenamefont {Raposo},\ and\ \citenamefont
  {Da~Luz}}]{viswanathan2008levy}%
  \BibitemOpen
  \bibfield  {author} {\bibinfo {author} {\bibfnamefont {G.~M.}\ \bibnamefont
  {Viswanathan}}, \bibinfo {author} {\bibfnamefont {E.}~\bibnamefont
  {Raposo}},\ and\ \bibinfo {author} {\bibfnamefont {M.}~\bibnamefont
  {Da~Luz}},\ }\bibfield  {title} {\bibinfo {title} {L{\'e}vy flights and
  superdiffusion in the context of biological encounters and random searches},\
  }\href@noop {} {\bibfield  {journal} {\bibinfo  {journal} {Physics of Life
  Reviews}\ }\textbf {\bibinfo {volume} {5}},\ \bibinfo {pages} {133} (\bibinfo
  {year} {2008})}\BibitemShut {NoStop}%
\bibitem [{\citenamefont {Slade}(1994)}]{slade1994self}%
  \BibitemOpen
  \bibfield  {author} {\bibinfo {author} {\bibfnamefont {G.}~\bibnamefont
  {Slade}},\ }\bibfield  {title} {\bibinfo {title} {Self-avoiding walks},\
  }\href@noop {} {\bibfield  {journal} {\bibinfo  {journal} {The Mathematical
  Intelligencer}\ }\textbf {\bibinfo {volume} {16}},\ \bibinfo {pages} {29}
  (\bibinfo {year} {1994})}\BibitemShut {NoStop}%
\bibitem [{\citenamefont {Slade}(2019)}]{slade2019self}%
  \BibitemOpen
  \bibfield  {author} {\bibinfo {author} {\bibfnamefont {G.}~\bibnamefont
  {Slade}},\ }\bibfield  {title} {\bibinfo {title} {Self-avoiding walk, spin
  systems and renormalization},\ }\href@noop {} {\bibfield  {journal} {\bibinfo
   {journal} {Proceedings of the Royal Society A}\ }\textbf {\bibinfo {volume}
  {475}},\ \bibinfo {pages} {20180549} (\bibinfo {year} {2019})}\BibitemShut
  {NoStop}%
\bibitem [{\citenamefont {Li}\ \emph {et~al.}(2011)\citenamefont {Li},
  \citenamefont {Cox},\ and\ \citenamefont {Flyvbjerg}}]{li2011dicty}%
  \BibitemOpen
  \bibfield  {author} {\bibinfo {author} {\bibfnamefont {L.}~\bibnamefont
  {Li}}, \bibinfo {author} {\bibfnamefont {E.~C.}\ \bibnamefont {Cox}},\ and\
  \bibinfo {author} {\bibfnamefont {H.}~\bibnamefont {Flyvbjerg}},\ }\bibfield
  {title} {\bibinfo {title} {‘\text{Dicty} dynamics’:
  \textit{Dictyostelium} motility as persistent random motion},\ }\href@noop {}
  {\bibfield  {journal} {\bibinfo  {journal} {Physical biology}\ }\textbf
  {\bibinfo {volume} {8}},\ \bibinfo {pages} {046006} (\bibinfo {year}
  {2011})}\BibitemShut {NoStop}%
\bibitem [{\citenamefont {Shaebani}\ and\ \citenamefont
  {Rieger}(2019)}]{shaebani2019transient}%
  \BibitemOpen
  \bibfield  {author} {\bibinfo {author} {\bibfnamefont {M.~R.}\ \bibnamefont
  {Shaebani}}\ and\ \bibinfo {author} {\bibfnamefont {H.}~\bibnamefont
  {Rieger}},\ }\bibfield  {title} {\bibinfo {title} {Transient anomalous
  diffusion in run-and-tumble dynamics},\ }\href@noop {} {\bibfield  {journal}
  {\bibinfo  {journal} {Frontiers in Physics}\ }\textbf {\bibinfo {volume}
  {7}},\ \bibinfo {pages} {120} (\bibinfo {year} {2019})}\BibitemShut {NoStop}%
\bibitem [{\citenamefont {Tero}\ \emph {et~al.}(2010)\citenamefont {Tero},
  \citenamefont {Takagi}, \citenamefont {Saigusa}, \citenamefont {Ito},
  \citenamefont {Bebber}, \citenamefont {Fricker}, \citenamefont {Yumiki},
  \citenamefont {Kobayashi},\ and\ \citenamefont {Nakagaki}}]{tero2010rules}%
  \BibitemOpen
  \bibfield  {author} {\bibinfo {author} {\bibfnamefont {A.}~\bibnamefont
  {Tero}}, \bibinfo {author} {\bibfnamefont {S.}~\bibnamefont {Takagi}},
  \bibinfo {author} {\bibfnamefont {T.}~\bibnamefont {Saigusa}}, \bibinfo
  {author} {\bibfnamefont {K.}~\bibnamefont {Ito}}, \bibinfo {author}
  {\bibfnamefont {D.~P.}\ \bibnamefont {Bebber}}, \bibinfo {author}
  {\bibfnamefont {M.~D.}\ \bibnamefont {Fricker}}, \bibinfo {author}
  {\bibfnamefont {K.}~\bibnamefont {Yumiki}}, \bibinfo {author} {\bibfnamefont
  {R.}~\bibnamefont {Kobayashi}},\ and\ \bibinfo {author} {\bibfnamefont
  {T.}~\bibnamefont {Nakagaki}},\ }\bibfield  {title} {\bibinfo {title} {Rules
  for biologically inspired adaptive network design},\ }\href@noop {}
  {\bibfield  {journal} {\bibinfo  {journal} {Science}\ }\textbf {\bibinfo
  {volume} {327}},\ \bibinfo {pages} {439} (\bibinfo {year}
  {2010})}\BibitemShut {NoStop}%
\bibitem [{\citenamefont {Latty}\ and\ \citenamefont
  {Beekman}(2011)}]{latty2011speed}%
  \BibitemOpen
  \bibfield  {author} {\bibinfo {author} {\bibfnamefont {T.}~\bibnamefont
  {Latty}}\ and\ \bibinfo {author} {\bibfnamefont {M.}~\bibnamefont
  {Beekman}},\ }\bibfield  {title} {\bibinfo {title} {Speed--accuracy
  trade-offs during foraging decisions in the acellular slime mould
  \textit{Physarum polycephalum}},\ }\href@noop {} {\bibfield  {journal}
  {\bibinfo  {journal} {Proceedings of the Royal Society B: Biological
  Sciences}\ }\textbf {\bibinfo {volume} {278}},\ \bibinfo {pages} {539}
  (\bibinfo {year} {2011})}\BibitemShut {NoStop}%
\bibitem [{\citenamefont {Dussutour}\ \emph {et~al.}(2010)\citenamefont
  {Dussutour}, \citenamefont {Latty}, \citenamefont {Beekman},\ and\
  \citenamefont {Simpson}}]{dussutour2010amoeboid}%
  \BibitemOpen
  \bibfield  {author} {\bibinfo {author} {\bibfnamefont {A.}~\bibnamefont
  {Dussutour}}, \bibinfo {author} {\bibfnamefont {T.}~\bibnamefont {Latty}},
  \bibinfo {author} {\bibfnamefont {M.}~\bibnamefont {Beekman}},\ and\ \bibinfo
  {author} {\bibfnamefont {S.~J.}\ \bibnamefont {Simpson}},\ }\bibfield
  {title} {\bibinfo {title} {Amoeboid organism solves complex nutritional
  challenges},\ }\href@noop {} {\bibfield  {journal} {\bibinfo  {journal}
  {Proceedings of the National Academy of Sciences}\ }\textbf {\bibinfo
  {volume} {107}},\ \bibinfo {pages} {4607} (\bibinfo {year}
  {2010})}\BibitemShut {NoStop}%
\bibitem [{\citenamefont {Verge-Serandour}\ and\ \citenamefont
  {Alim}(2023)}]{verge2023physarum}%
  \BibitemOpen
  \bibfield  {author} {\bibinfo {author} {\bibfnamefont {M.~L.}\ \bibnamefont
  {Verge-Serandour}}\ and\ \bibinfo {author} {\bibfnamefont {K.}~\bibnamefont
  {Alim}},\ }\bibfield  {title} {\bibinfo {title} {\textit{Physarum
  polycephalum}: Smart network adaptation},\ }\href@noop {} {\bibfield
  {journal} {\bibinfo  {journal} {arXiv preprint arXiv:2306.09063}\ } (\bibinfo
  {year} {2023})}\BibitemShut {NoStop}%
\bibitem [{\citenamefont {Reid}\ \emph {et~al.}(2012)\citenamefont {Reid},
  \citenamefont {Latty}, \citenamefont {Dussutour},\ and\ \citenamefont
  {Beekman}}]{reid2012slime}%
  \BibitemOpen
  \bibfield  {author} {\bibinfo {author} {\bibfnamefont {C.~R.}\ \bibnamefont
  {Reid}}, \bibinfo {author} {\bibfnamefont {T.}~\bibnamefont {Latty}},
  \bibinfo {author} {\bibfnamefont {A.}~\bibnamefont {Dussutour}},\ and\
  \bibinfo {author} {\bibfnamefont {M.}~\bibnamefont {Beekman}},\ }\bibfield
  {title} {\bibinfo {title} {Slime mold uses an externalized spatial
  “memory” to navigate in complex environments},\ }\href@noop {} {\bibfield
   {journal} {\bibinfo  {journal} {Proceedings of the National Academy of
  Sciences}\ }\textbf {\bibinfo {volume} {109}},\ \bibinfo {pages} {17490}
  (\bibinfo {year} {2012})}\BibitemShut {NoStop}%
\bibitem [{\citenamefont {Reid}\ \emph {et~al.}(2013)\citenamefont {Reid},
  \citenamefont {Beekman}, \citenamefont {Latty},\ and\ \citenamefont
  {Dussutour}}]{reid2013amoeboid}%
  \BibitemOpen
  \bibfield  {author} {\bibinfo {author} {\bibfnamefont {C.~R.}\ \bibnamefont
  {Reid}}, \bibinfo {author} {\bibfnamefont {M.}~\bibnamefont {Beekman}},
  \bibinfo {author} {\bibfnamefont {T.}~\bibnamefont {Latty}},\ and\ \bibinfo
  {author} {\bibfnamefont {A.}~\bibnamefont {Dussutour}},\ }\bibfield  {title}
  {\bibinfo {title} {Amoeboid organism uses extracellular secretions to make
  smart foraging decisions},\ }\href@noop {} {\bibfield  {journal} {\bibinfo
  {journal} {Behavioral Ecology}\ }\textbf {\bibinfo {volume} {24}},\ \bibinfo
  {pages} {812} (\bibinfo {year} {2013})}\BibitemShut {NoStop}%
\bibitem [{\citenamefont {Smith-Ferguson}\ \emph {et~al.}(2017)\citenamefont
  {Smith-Ferguson}, \citenamefont {Reid}, \citenamefont {Latty},\ and\
  \citenamefont {Beekman}}]{smith2017hansel}%
  \BibitemOpen
  \bibfield  {author} {\bibinfo {author} {\bibfnamefont {J.}~\bibnamefont
  {Smith-Ferguson}}, \bibinfo {author} {\bibfnamefont {C.~R.}\ \bibnamefont
  {Reid}}, \bibinfo {author} {\bibfnamefont {T.}~\bibnamefont {Latty}},\ and\
  \bibinfo {author} {\bibfnamefont {M.}~\bibnamefont {Beekman}},\ }\bibfield
  {title} {\bibinfo {title} {H{\"a}nsel, gretel and the slime mould—how an
  external spatial memory aids navigation in complex environments},\
  }\href@noop {} {\bibfield  {journal} {\bibinfo  {journal} {Journal of Physics
  D: Applied Physics}\ }\textbf {\bibinfo {volume} {50}},\ \bibinfo {pages}
  {414003} (\bibinfo {year} {2017})}\BibitemShut {NoStop}%
\bibitem [{\citenamefont {Kuroda}\ \emph {et~al.}(2015)\citenamefont {Kuroda},
  \citenamefont {Takagi}, \citenamefont {Nakagaki},\ and\ \citenamefont
  {Ueda}}]{kuroda2015allometry}%
  \BibitemOpen
  \bibfield  {author} {\bibinfo {author} {\bibfnamefont {S.}~\bibnamefont
  {Kuroda}}, \bibinfo {author} {\bibfnamefont {S.}~\bibnamefont {Takagi}},
  \bibinfo {author} {\bibfnamefont {T.}~\bibnamefont {Nakagaki}},\ and\
  \bibinfo {author} {\bibfnamefont {T.}~\bibnamefont {Ueda}},\ }\bibfield
  {title} {\bibinfo {title} {Allometry in \textit{Physarum} plasmodium during
  free locomotion: Size versus shape, speed and rhythm},\ }\href@noop {}
  {\bibfield  {journal} {\bibinfo  {journal} {Journal of experimental biology}\
  }\textbf {\bibinfo {volume} {218}},\ \bibinfo {pages} {3729} (\bibinfo {year}
  {2015})}\BibitemShut {NoStop}%
\bibitem [{\citenamefont {Viswanathan}\ \emph {et~al.}(2005)\citenamefont
  {Viswanathan}, \citenamefont {Raposo}, \citenamefont {Bartumeus},
  \citenamefont {Catalan},\ and\ \citenamefont
  {Da~Luz}}]{viswanathan2005necessary}%
  \BibitemOpen
  \bibfield  {author} {\bibinfo {author} {\bibfnamefont {G.}~\bibnamefont
  {Viswanathan}}, \bibinfo {author} {\bibfnamefont {E.}~\bibnamefont {Raposo}},
  \bibinfo {author} {\bibfnamefont {F.}~\bibnamefont {Bartumeus}}, \bibinfo
  {author} {\bibfnamefont {J.}~\bibnamefont {Catalan}},\ and\ \bibinfo {author}
  {\bibfnamefont {M.}~\bibnamefont {Da~Luz}},\ }\bibfield  {title} {\bibinfo
  {title} {Necessary criterion for distinguishing true superdiffusion from
  correlated random walk processes},\ }\href@noop {} {\bibfield  {journal}
  {\bibinfo  {journal} {Physical Review E}\ }\textbf {\bibinfo {volume} {72}},\
  \bibinfo {pages} {011111} (\bibinfo {year} {2005})}\BibitemShut {NoStop}%
\bibitem [{\citenamefont {Rodiek}\ \emph {et~al.}(2015)\citenamefont {Rodiek},
  \citenamefont {Takagi}, \citenamefont {Ueda},\ and\ \citenamefont
  {Hauser}}]{rodiek2015patterns}%
  \BibitemOpen
  \bibfield  {author} {\bibinfo {author} {\bibfnamefont {B.}~\bibnamefont
  {Rodiek}}, \bibinfo {author} {\bibfnamefont {S.}~\bibnamefont {Takagi}},
  \bibinfo {author} {\bibfnamefont {T.}~\bibnamefont {Ueda}},\ and\ \bibinfo
  {author} {\bibfnamefont {M.~J.}\ \bibnamefont {Hauser}},\ }\bibfield  {title}
  {\bibinfo {title} {Patterns of cell thickness oscillations during directional
  migration of \textit{Physarum polycephalum}},\ }\href@noop {} {\bibfield
  {journal} {\bibinfo  {journal} {European Biophysics Journal}\ }\textbf
  {\bibinfo {volume} {44}},\ \bibinfo {pages} {349} (\bibinfo {year}
  {2015})}\BibitemShut {NoStop}%
\bibitem [{\citenamefont {B{\'e}nichou}\ \emph {et~al.}(2011)\citenamefont
  {B{\'e}nichou}, \citenamefont {Loverdo}, \citenamefont {Moreau},\ and\
  \citenamefont {Voituriez}}]{benichou2011intermittent}%
  \BibitemOpen
  \bibfield  {author} {\bibinfo {author} {\bibfnamefont {O.}~\bibnamefont
  {B{\'e}nichou}}, \bibinfo {author} {\bibfnamefont {C.}~\bibnamefont
  {Loverdo}}, \bibinfo {author} {\bibfnamefont {M.}~\bibnamefont {Moreau}},\
  and\ \bibinfo {author} {\bibfnamefont {R.}~\bibnamefont {Voituriez}},\
  }\bibfield  {title} {\bibinfo {title} {Intermittent search strategies},\
  }\href@noop {} {\bibfield  {journal} {\bibinfo  {journal} {Reviews of Modern
  Physics}\ }\textbf {\bibinfo {volume} {83}},\ \bibinfo {pages} {81} (\bibinfo
  {year} {2011})}\BibitemShut {NoStop}%
\bibitem [{\citenamefont {Varennes}\ \emph {et~al.}(2019)\citenamefont
  {Varennes}, \citenamefont {Moon}, \citenamefont {Saha}, \citenamefont
  {Mugler},\ and\ \citenamefont {Han}}]{varennes2019physical}%
  \BibitemOpen
  \bibfield  {author} {\bibinfo {author} {\bibfnamefont {J.}~\bibnamefont
  {Varennes}}, \bibinfo {author} {\bibfnamefont {H.-r.}\ \bibnamefont {Moon}},
  \bibinfo {author} {\bibfnamefont {S.}~\bibnamefont {Saha}}, \bibinfo {author}
  {\bibfnamefont {A.}~\bibnamefont {Mugler}},\ and\ \bibinfo {author}
  {\bibfnamefont {B.}~\bibnamefont {Han}},\ }\bibfield  {title} {\bibinfo
  {title} {Physical constraints on accuracy and persistence during breast
  cancer cell chemotaxis},\ }\href@noop {} {\bibfield  {journal} {\bibinfo
  {journal} {PLOS Computational Biology}\ }\textbf {\bibinfo {volume} {15}},\
  \bibinfo {pages} {e1006961} (\bibinfo {year} {2019})}\BibitemShut {NoStop}%
\bibitem [{\citenamefont {Hemmer}\ and\ \citenamefont
  {Hemmer}(1986)}]{hemmer1986trapping}%
  \BibitemOpen
  \bibfield  {author} {\bibinfo {author} {\bibfnamefont {P.}~\bibnamefont
  {Hemmer}}\ and\ \bibinfo {author} {\bibfnamefont {S.}~\bibnamefont
  {Hemmer}},\ }\bibfield  {title} {\bibinfo {title} {Trapping of genuine
  self-avoiding walks},\ }\href@noop {} {\bibfield  {journal} {\bibinfo
  {journal} {Physical Review A}\ }\textbf {\bibinfo {volume} {34}},\ \bibinfo
  {pages} {3304} (\bibinfo {year} {1986})}\BibitemShut {NoStop}%
\bibitem [{\citenamefont {Zhao}\ \emph {et~al.}(2013)\citenamefont {Zhao},
  \citenamefont {Tseng}, \citenamefont {Beckerman}, \citenamefont {Jin},
  \citenamefont {Gibiansky}, \citenamefont {Harrison}, \citenamefont {Luijten},
  \citenamefont {Parsek},\ and\ \citenamefont {Wong}}]{zhao2013psl}%
  \BibitemOpen
  \bibfield  {author} {\bibinfo {author} {\bibfnamefont {K.}~\bibnamefont
  {Zhao}}, \bibinfo {author} {\bibfnamefont {B.~S.}\ \bibnamefont {Tseng}},
  \bibinfo {author} {\bibfnamefont {B.}~\bibnamefont {Beckerman}}, \bibinfo
  {author} {\bibfnamefont {F.}~\bibnamefont {Jin}}, \bibinfo {author}
  {\bibfnamefont {M.~L.}\ \bibnamefont {Gibiansky}}, \bibinfo {author}
  {\bibfnamefont {J.~J.}\ \bibnamefont {Harrison}}, \bibinfo {author}
  {\bibfnamefont {E.}~\bibnamefont {Luijten}}, \bibinfo {author} {\bibfnamefont
  {M.~R.}\ \bibnamefont {Parsek}},\ and\ \bibinfo {author} {\bibfnamefont
  {G.~C.}\ \bibnamefont {Wong}},\ }\bibfield  {title} {\bibinfo {title} {Psl
  trails guide exploration and microcolony formation in \textit{Pseudomonas
  aeruginosa} biofilms},\ }\href@noop {} {\bibfield  {journal} {\bibinfo
  {journal} {Nature}\ }\textbf {\bibinfo {volume} {497}},\ \bibinfo {pages}
  {388} (\bibinfo {year} {2013})}\BibitemShut {NoStop}%
\bibitem [{\citenamefont {D’alessandro}\ \emph {et~al.}(2021)\citenamefont
  {D’alessandro}, \citenamefont {Barbier-Chebbah}, \citenamefont {Cellerin},
  \citenamefont {Benichou}, \citenamefont {M{\`e}ge}, \citenamefont
  {Voituriez},\ and\ \citenamefont {Ladoux}}]{d2021cell}%
  \BibitemOpen
  \bibfield  {author} {\bibinfo {author} {\bibfnamefont {J.}~\bibnamefont
  {D’alessandro}}, \bibinfo {author} {\bibfnamefont {A.}~\bibnamefont
  {Barbier-Chebbah}}, \bibinfo {author} {\bibfnamefont {V.}~\bibnamefont
  {Cellerin}}, \bibinfo {author} {\bibfnamefont {O.}~\bibnamefont {Benichou}},
  \bibinfo {author} {\bibfnamefont {R.~M.}\ \bibnamefont {M{\`e}ge}}, \bibinfo
  {author} {\bibfnamefont {R.}~\bibnamefont {Voituriez}},\ and\ \bibinfo
  {author} {\bibfnamefont {B.}~\bibnamefont {Ladoux}},\ }\bibfield  {title}
  {\bibinfo {title} {Cell migration guided by long-lived spatial memory},\
  }\href@noop {} {\bibfield  {journal} {\bibinfo  {journal} {Nature
  Communications}\ }\textbf {\bibinfo {volume} {12}},\ \bibinfo {pages} {4118}
  (\bibinfo {year} {2021})}\BibitemShut {NoStop}%
\bibitem [{\citenamefont {Robinson}\ \emph {et~al.}(2005)\citenamefont
  {Robinson}, \citenamefont {Jackson}, \citenamefont {Holcombe},\ and\
  \citenamefont {Ratnieks}}]{robinson2005no}%
  \BibitemOpen
  \bibfield  {author} {\bibinfo {author} {\bibfnamefont {E.~J.}\ \bibnamefont
  {Robinson}}, \bibinfo {author} {\bibfnamefont {D.~E.}\ \bibnamefont
  {Jackson}}, \bibinfo {author} {\bibfnamefont {M.}~\bibnamefont {Holcombe}},\
  and\ \bibinfo {author} {\bibfnamefont {F.~L.}\ \bibnamefont {Ratnieks}},\
  }\bibfield  {title} {\bibinfo {title} {‘no entry’signal in ant
  foraging},\ }\href@noop {} {\bibfield  {journal} {\bibinfo  {journal}
  {Nature}\ }\textbf {\bibinfo {volume} {438}},\ \bibinfo {pages} {442}
  (\bibinfo {year} {2005})}\BibitemShut {NoStop}%
\bibitem [{\citenamefont {Kranz}\ \emph {et~al.}(2016)\citenamefont {Kranz},
  \citenamefont {Gelimson}, \citenamefont {Zhao}, \citenamefont {Wong},\ and\
  \citenamefont {Golestanian}}]{kranz2016effective}%
  \BibitemOpen
  \bibfield  {author} {\bibinfo {author} {\bibfnamefont {W.~T.}\ \bibnamefont
  {Kranz}}, \bibinfo {author} {\bibfnamefont {A.}~\bibnamefont {Gelimson}},
  \bibinfo {author} {\bibfnamefont {K.}~\bibnamefont {Zhao}}, \bibinfo {author}
  {\bibfnamefont {G.~C.}\ \bibnamefont {Wong}},\ and\ \bibinfo {author}
  {\bibfnamefont {R.}~\bibnamefont {Golestanian}},\ }\bibfield  {title}
  {\bibinfo {title} {Effective dynamics of microorganisms that interact with
  their own trail},\ }\href@noop {} {\bibfield  {journal} {\bibinfo  {journal}
  {Physical review letters}\ }\textbf {\bibinfo {volume} {117}},\ \bibinfo
  {pages} {038101} (\bibinfo {year} {2016})}\BibitemShut {NoStop}%
\bibitem [{\citenamefont {Barbier-Chebbah}\ \emph {et~al.}(2022)\citenamefont
  {Barbier-Chebbah}, \citenamefont {B{\'e}nichou},\ and\ \citenamefont
  {Voituriez}}]{barbier2022self}%
  \BibitemOpen
  \bibfield  {author} {\bibinfo {author} {\bibfnamefont {A.}~\bibnamefont
  {Barbier-Chebbah}}, \bibinfo {author} {\bibfnamefont {O.}~\bibnamefont
  {B{\'e}nichou}},\ and\ \bibinfo {author} {\bibfnamefont {R.}~\bibnamefont
  {Voituriez}},\ }\bibfield  {title} {\bibinfo {title} {Self-interacting random
  walks: Aging, exploration, and first-passage times},\ }\href@noop {}
  {\bibfield  {journal} {\bibinfo  {journal} {Physical Review X}\ }\textbf
  {\bibinfo {volume} {12}},\ \bibinfo {pages} {011052} (\bibinfo {year}
  {2022})}\BibitemShut {NoStop}%
\bibitem [{\citenamefont {Moreau}\ \emph {et~al.}(2009)\citenamefont {Moreau},
  \citenamefont {B{\'e}nichou}, \citenamefont {Loverdo},\ and\ \citenamefont
  {Voituriez}}]{moreau2009chance}%
  \BibitemOpen
  \bibfield  {author} {\bibinfo {author} {\bibfnamefont {M.}~\bibnamefont
  {Moreau}}, \bibinfo {author} {\bibfnamefont {O.}~\bibnamefont
  {B{\'e}nichou}}, \bibinfo {author} {\bibfnamefont {C.}~\bibnamefont
  {Loverdo}},\ and\ \bibinfo {author} {\bibfnamefont {R.}~\bibnamefont
  {Voituriez}},\ }\bibfield  {title} {\bibinfo {title} {Chance and strategy in
  search processes},\ }\href@noop {} {\bibfield  {journal} {\bibinfo  {journal}
  {Journal of Statistical Mechanics: Theory and Experiment}\ }\textbf {\bibinfo
  {volume} {2009}},\ \bibinfo {pages} {P12006} (\bibinfo {year}
  {2009})}\BibitemShut {NoStop}%
\bibitem [{\citenamefont {Sims}\ \emph {et~al.}(2014)\citenamefont {Sims},
  \citenamefont {Reynolds}, \citenamefont {Humphries}, \citenamefont
  {Southall}, \citenamefont {Wearmouth}, \citenamefont {Metcalfe},\ and\
  \citenamefont {Twitchett}}]{sims2014hierarchical}%
  \BibitemOpen
  \bibfield  {author} {\bibinfo {author} {\bibfnamefont {D.~W.}\ \bibnamefont
  {Sims}}, \bibinfo {author} {\bibfnamefont {A.~M.}\ \bibnamefont {Reynolds}},
  \bibinfo {author} {\bibfnamefont {N.~E.}\ \bibnamefont {Humphries}}, \bibinfo
  {author} {\bibfnamefont {E.~J.}\ \bibnamefont {Southall}}, \bibinfo {author}
  {\bibfnamefont {V.~J.}\ \bibnamefont {Wearmouth}}, \bibinfo {author}
  {\bibfnamefont {B.}~\bibnamefont {Metcalfe}},\ and\ \bibinfo {author}
  {\bibfnamefont {R.~J.}\ \bibnamefont {Twitchett}},\ }\bibfield  {title}
  {\bibinfo {title} {Hierarchical random walks in trace fossils and the origin
  of optimal search behavior},\ }\href@noop {} {\bibfield  {journal} {\bibinfo
  {journal} {Proceedings of the National Academy of Sciences}\ }\textbf
  {\bibinfo {volume} {111}},\ \bibinfo {pages} {11073} (\bibinfo {year}
  {2014})}\BibitemShut {NoStop}%
\bibitem [{\citenamefont {Barbara}\ and\ \citenamefont
  {Mitchell}(2003)}]{barbara2003marine}%
  \BibitemOpen
  \bibfield  {author} {\bibinfo {author} {\bibfnamefont {G.~M.}\ \bibnamefont
  {Barbara}}\ and\ \bibinfo {author} {\bibfnamefont {J.~G.}\ \bibnamefont
  {Mitchell}},\ }\bibfield  {title} {\bibinfo {title} {Marine bacterial
  organisation around point-like sources of amino acids},\ }\href@noop {}
  {\bibfield  {journal} {\bibinfo  {journal} {FEMS microbiology ecology}\
  }\textbf {\bibinfo {volume} {43}},\ \bibinfo {pages} {99} (\bibinfo {year}
  {2003})}\BibitemShut {NoStop}%
\bibitem [{\citenamefont {Bartumeus}\ \emph {et~al.}(2003)\citenamefont
  {Bartumeus}, \citenamefont {Peters}, \citenamefont {Pueyo}, \citenamefont
  {Marras{\'e}},\ and\ \citenamefont {Catalan}}]{bartumeus2003helical}%
  \BibitemOpen
  \bibfield  {author} {\bibinfo {author} {\bibfnamefont {F.}~\bibnamefont
  {Bartumeus}}, \bibinfo {author} {\bibfnamefont {F.}~\bibnamefont {Peters}},
  \bibinfo {author} {\bibfnamefont {S.}~\bibnamefont {Pueyo}}, \bibinfo
  {author} {\bibfnamefont {C.}~\bibnamefont {Marras{\'e}}},\ and\ \bibinfo
  {author} {\bibfnamefont {J.}~\bibnamefont {Catalan}},\ }\bibfield  {title}
  {\bibinfo {title} {Helical \text{L{\'e}vy} walks: adjusting searching
  statistics to resource availability in microzooplankton},\ }\href@noop {}
  {\bibfield  {journal} {\bibinfo  {journal} {Proceedings of the National
  Academy of Sciences}\ }\textbf {\bibinfo {volume} {100}},\ \bibinfo {pages}
  {12771} (\bibinfo {year} {2003})}\BibitemShut {NoStop}%
\bibitem [{\citenamefont {Varnum}\ and\ \citenamefont
  {Soll}(1984)}]{varnum1984effects}%
  \BibitemOpen
  \bibfield  {author} {\bibinfo {author} {\bibfnamefont {B.}~\bibnamefont
  {Varnum}}\ and\ \bibinfo {author} {\bibfnamefont {D.~R.}\ \bibnamefont
  {Soll}},\ }\bibfield  {title} {\bibinfo {title} {Effects of \text{cAMP} on
  single cell motility in \textit{Dictyostelium}.},\ }\href@noop {} {\bibfield
  {journal} {\bibinfo  {journal} {The Journal of cell biology}\ }\textbf
  {\bibinfo {volume} {99}},\ \bibinfo {pages} {1151} (\bibinfo {year}
  {1984})}\BibitemShut {NoStop}%
\bibitem [{\citenamefont {Tweedy}\ \emph {et~al.}(2020)\citenamefont {Tweedy},
  \citenamefont {Thomason}, \citenamefont {Paschke}, \citenamefont {Martin},
  \citenamefont {Machesky}, \citenamefont {Zagnoni},\ and\ \citenamefont
  {Insall}}]{tweedy2020seeing}%
  \BibitemOpen
  \bibfield  {author} {\bibinfo {author} {\bibfnamefont {L.}~\bibnamefont
  {Tweedy}}, \bibinfo {author} {\bibfnamefont {P.~A.}\ \bibnamefont
  {Thomason}}, \bibinfo {author} {\bibfnamefont {P.~I.}\ \bibnamefont
  {Paschke}}, \bibinfo {author} {\bibfnamefont {K.}~\bibnamefont {Martin}},
  \bibinfo {author} {\bibfnamefont {L.~M.}\ \bibnamefont {Machesky}}, \bibinfo
  {author} {\bibfnamefont {M.}~\bibnamefont {Zagnoni}},\ and\ \bibinfo {author}
  {\bibfnamefont {R.~H.}\ \bibnamefont {Insall}},\ }\bibfield  {title}
  {\bibinfo {title} {Seeing around corners: Cells solve mazes and respond at a
  distance using attractant breakdown},\ }\href@noop {} {\bibfield  {journal}
  {\bibinfo  {journal} {Science}\ }\textbf {\bibinfo {volume} {369}},\ \bibinfo
  {pages} {eaay9792} (\bibinfo {year} {2020})}\BibitemShut {NoStop}%
\bibitem [{\citenamefont {Gerber}\ \emph {et~al.}(2022)\citenamefont {Gerber},
  \citenamefont {Loureiro}, \citenamefont {Schramma}, \citenamefont {Chen},
  \citenamefont {Jain}, \citenamefont {Weber}, \citenamefont {Weigert},
  \citenamefont {Santel}, \citenamefont {Alim}, \citenamefont {Treutlein} \emph
  {et~al.}}]{gerber2022spatial}%
  \BibitemOpen
  \bibfield  {author} {\bibinfo {author} {\bibfnamefont {T.}~\bibnamefont
  {Gerber}}, \bibinfo {author} {\bibfnamefont {C.}~\bibnamefont {Loureiro}},
  \bibinfo {author} {\bibfnamefont {N.}~\bibnamefont {Schramma}}, \bibinfo
  {author} {\bibfnamefont {S.}~\bibnamefont {Chen}}, \bibinfo {author}
  {\bibfnamefont {A.}~\bibnamefont {Jain}}, \bibinfo {author} {\bibfnamefont
  {A.}~\bibnamefont {Weber}}, \bibinfo {author} {\bibfnamefont
  {A.}~\bibnamefont {Weigert}}, \bibinfo {author} {\bibfnamefont
  {M.}~\bibnamefont {Santel}}, \bibinfo {author} {\bibfnamefont
  {K.}~\bibnamefont {Alim}}, \bibinfo {author} {\bibfnamefont {B.}~\bibnamefont
  {Treutlein}}, \emph {et~al.},\ }\bibfield  {title} {\bibinfo {title} {Spatial
  transcriptomic and single-nucleus analysis reveals heterogeneity in a
  gigantic single-celled syncytium},\ }\href@noop {} {\bibfield  {journal}
  {\bibinfo  {journal} {Elife}\ }\textbf {\bibinfo {volume} {11}},\ \bibinfo
  {pages} {e69745} (\bibinfo {year} {2022})}\BibitemShut {NoStop}%
\bibitem [{\citenamefont {Daniel}\ and\ \citenamefont
  {Rusch}(1961)}]{daniel1961pure}%
  \BibitemOpen
  \bibfield  {author} {\bibinfo {author} {\bibfnamefont {J.~W.}\ \bibnamefont
  {Daniel}}\ and\ \bibinfo {author} {\bibfnamefont {H.~P.}\ \bibnamefont
  {Rusch}},\ }\bibfield  {title} {\bibinfo {title} {The pure culture of
  \textit{Physarum polycephalum} on a partially defined soluble medium},\
  }\href@noop {} {\bibfield  {journal} {\bibinfo  {journal} {Microbiology}\
  }\textbf {\bibinfo {volume} {25}},\ \bibinfo {pages} {47} (\bibinfo {year}
  {1961})}\BibitemShut {NoStop}%
\bibitem [{\citenamefont {Daniel}\ \emph {et~al.}(1962)\citenamefont {Daniel},
  \citenamefont {Kelley},\ and\ \citenamefont {Rusch}}]{daniel1962hematin}%
  \BibitemOpen
  \bibfield  {author} {\bibinfo {author} {\bibfnamefont {J.~W.}\ \bibnamefont
  {Daniel}}, \bibinfo {author} {\bibfnamefont {J.}~\bibnamefont {Kelley}},\
  and\ \bibinfo {author} {\bibfnamefont {H.~P.}\ \bibnamefont {Rusch}},\
  }\bibfield  {title} {\bibinfo {title} {Hematin-requiring plasmodial
  myxomycete},\ }\href@noop {} {\bibfield  {journal} {\bibinfo  {journal}
  {Journal of Bacteriology}\ }\textbf {\bibinfo {volume} {84}},\ \bibinfo
  {pages} {1104} (\bibinfo {year} {1962})}\BibitemShut {NoStop}%
\bibitem [{\citenamefont {Hato}\ \emph {et~al.}(1976)\citenamefont {Hato},
  \citenamefont {Ueda}, \citenamefont {Kurihara},\ and\ \citenamefont
  {Kobatake}}]{hato1976phototaxis}%
  \BibitemOpen
  \bibfield  {author} {\bibinfo {author} {\bibfnamefont {M.}~\bibnamefont
  {Hato}}, \bibinfo {author} {\bibfnamefont {T.}~\bibnamefont {Ueda}}, \bibinfo
  {author} {\bibfnamefont {K.}~\bibnamefont {Kurihara}},\ and\ \bibinfo
  {author} {\bibfnamefont {Y.}~\bibnamefont {Kobatake}},\ }\bibfield  {title}
  {\bibinfo {title} {Phototaxis in true slime mold \textit{Physarum
  polycephalum}},\ }\href@noop {} {\bibfield  {journal} {\bibinfo  {journal}
  {Cell Structure and function}\ }\textbf {\bibinfo {volume} {1}},\ \bibinfo
  {pages} {269} (\bibinfo {year} {1976})}\BibitemShut {NoStop}%
\bibitem [{\citenamefont {Blattner}\ \emph {et~al.}(2014)\citenamefont
  {Blattner}, \citenamefont {Keyrouz}, \citenamefont {Chalfoun}, \citenamefont
  {Stivalet}, \citenamefont {Brady},\ and\ \citenamefont
  {Zhou}}]{blattner2014hybrid}%
  \BibitemOpen
  \bibfield  {author} {\bibinfo {author} {\bibfnamefont {T.}~\bibnamefont
  {Blattner}}, \bibinfo {author} {\bibfnamefont {W.}~\bibnamefont {Keyrouz}},
  \bibinfo {author} {\bibfnamefont {J.}~\bibnamefont {Chalfoun}}, \bibinfo
  {author} {\bibfnamefont {B.}~\bibnamefont {Stivalet}}, \bibinfo {author}
  {\bibfnamefont {M.}~\bibnamefont {Brady}},\ and\ \bibinfo {author}
  {\bibfnamefont {S.}~\bibnamefont {Zhou}},\ }\bibfield  {title} {\bibinfo
  {title} {A hybrid cpu-gpu system for stitching large scale optical microscopy
  images},\ }in\ \href@noop {} {\emph {\bibinfo {booktitle} {2014 43rd
  International Conference on Parallel Processing}}}\ (\bibinfo {organization}
  {IEEE},\ \bibinfo {year} {2014})\ pp.\ \bibinfo {pages} {1--9}\BibitemShut
  {NoStop}%
\bibitem [{\citenamefont {Chalfoun}\ \emph {et~al.}(2017)\citenamefont
  {Chalfoun}, \citenamefont {Majurski}, \citenamefont {Blattner}, \citenamefont
  {Bhadriraju}, \citenamefont {Keyrouz}, \citenamefont {Bajcsy},\ and\
  \citenamefont {Brady}}]{chalfoun2017mist}%
  \BibitemOpen
  \bibfield  {author} {\bibinfo {author} {\bibfnamefont {J.}~\bibnamefont
  {Chalfoun}}, \bibinfo {author} {\bibfnamefont {M.}~\bibnamefont {Majurski}},
  \bibinfo {author} {\bibfnamefont {T.}~\bibnamefont {Blattner}}, \bibinfo
  {author} {\bibfnamefont {K.}~\bibnamefont {Bhadriraju}}, \bibinfo {author}
  {\bibfnamefont {W.}~\bibnamefont {Keyrouz}}, \bibinfo {author} {\bibfnamefont
  {P.}~\bibnamefont {Bajcsy}},\ and\ \bibinfo {author} {\bibfnamefont
  {M.}~\bibnamefont {Brady}},\ }\bibfield  {title} {\bibinfo {title}
  {\text{MIST}: Accurate and scalable microscopy image stitching tool with
  stage modeling and error minimization},\ }\href@noop {} {\bibfield  {journal}
  {\bibinfo  {journal} {Scientific reports}\ }\textbf {\bibinfo {volume} {7}},\
  \bibinfo {pages} {4988} (\bibinfo {year} {2017})}\BibitemShut {NoStop}%
\end{thebibliography}%


\begin{thebibliography}{6}%
\makeatletter
\providecommand \@ifxundefined [1]{%
 \@ifx{#1\undefined}
}%
\providecommand \@ifnum [1]{%
 \ifnum #1\expandafter \@firstoftwo
 \else \expandafter \@secondoftwo
 \fi
}%
\providecommand \@ifx [1]{%
 \ifx #1\expandafter \@firstoftwo
 \else \expandafter \@secondoftwo
 \fi
}%
\providecommand \natexlab [1]{#1}%
\providecommand \enquote  [1]{``#1''}%
\providecommand \bibnamefont  [1]{#1}%
\providecommand \bibfnamefont [1]{#1}%
\providecommand \citenamefont [1]{#1}%
\providecommand \href@noop [0]{\@secondoftwo}%
\providecommand \href [0]{\begingroup \@sanitize@url \@href}%
\providecommand \@href[1]{\@@startlink{#1}\@@href}%
\providecommand \@@href[1]{\endgroup#1\@@endlink}%
\providecommand \@sanitize@url [0]{\catcode `\\12\catcode `\$12\catcode
  `\&12\catcode `\#12\catcode `\^12\catcode `\_12\catcode `\%12\relax}%
\providecommand \@@startlink[1]{}%
\providecommand \@@endlink[0]{}%
\providecommand \url  [0]{\begingroup\@sanitize@url \@url }%
\providecommand \@url [1]{\endgroup\@href {#1}{\urlprefix }}%
\providecommand \urlprefix  [0]{URL }%
\providecommand \Eprint [0]{\href }%
\providecommand \doibase [0]{https://doi.org/}%
\providecommand \selectlanguage [0]{\@gobble}%
\providecommand \bibinfo  [0]{\@secondoftwo}%
\providecommand \bibfield  [0]{\@secondoftwo}%
\providecommand \translation [1]{[#1]}%
\providecommand \BibitemOpen [0]{}%
\providecommand \bibitemStop [0]{}%
\providecommand \bibitemNoStop [0]{.\EOS\space}%
\providecommand \EOS [0]{\spacefactor3000\relax}%
\providecommand \BibitemShut  [1]{\csname bibitem#1\endcsname}%
\let\auto@bib@innerbib\@empty
\bibitem [{\citenamefont {Anderson}\ and\ \citenamefont
  {Burnham}(2004)}]{anderson2004model}%
  \BibitemOpen
  \bibfield  {author} {\bibinfo {author} {\bibnamefont {Anderson},
  \bibfnamefont {D.}}and\ \bibinfo {author} {\bibnamefont {Burnham},
  \bibfnamefont {K.}},\ }\bibfield  {title} {\enquote {\bibinfo {title} {Model
  selection and multi-model inference},}\ }\href@noop {} {\bibfield  {journal}
  {\bibinfo  {journal} {Second. NY: Springer-Verlag}\ }\textbf {\bibinfo
  {volume} {63}},\ \bibinfo {pages} {10} (\bibinfo {year} {2004})}\BibitemShut
  {NoStop}%
\bibitem [{\citenamefont {Devroye}(1986)}]{Devroye1986}%
  \BibitemOpen
  \bibfield  {author} {\bibinfo {author} {\bibnamefont {Devroye}, \bibfnamefont
  {L.}},\ }\enquote {\bibinfo {title} {General principles in random variate
  generation},}\ in\ \href {https://doi.org/10.1007/978-1-4613-8643-8_2} {\emph
  {\bibinfo {booktitle} {Non-Uniform Random Variate Generation}}}\ (\bibinfo
  {publisher} {Springer New York},\ \bibinfo {address} {New York, NY},\
  \bibinfo {year} {1986})\ pp.\ \bibinfo {pages} {27--82}\BibitemShut {NoStop}%
\bibitem [{\citenamefont {Hemmer}\ and\ \citenamefont
  {Hemmer}(1986)}]{hemmer1986trapping}%
  \BibitemOpen
  \bibfield  {author} {\bibinfo {author} {\bibnamefont {Hemmer}, \bibfnamefont
  {P.}}and\ \bibinfo {author} {\bibnamefont {Hemmer}, \bibfnamefont {S.}},\
  }\bibfield  {title} {\enquote {\bibinfo {title} {Trapping of genuine
  self-avoiding walks},}\ }\href@noop {} {\bibfield  {journal} {\bibinfo
  {journal} {Physical Review A}\ }\textbf {\bibinfo {volume} {34}},\ \bibinfo
  {pages} {3304} (\bibinfo {year} {1986})}\BibitemShut {NoStop}%
\bibitem [{\citenamefont {Lyklema}\ and\ \citenamefont
  {Kremer}(1986)}]{lyklema1986monte}%
  \BibitemOpen
  \bibfield  {author} {\bibinfo {author} {\bibnamefont {Lyklema}, \bibfnamefont
  {J.}}and\ \bibinfo {author} {\bibnamefont {Kremer}, \bibfnamefont {K.}},\
  }\bibfield  {title} {\enquote {\bibinfo {title} {Monte carlo series analysis
  of irreversible self-avoiding walks. ii. the growing self-avoiding walk},}\
  }\href@noop {} {\bibfield  {journal} {\bibinfo  {journal} {Journal of Physics
  A: Mathematical and General}\ }\textbf {\bibinfo {volume} {19}},\ \bibinfo
  {pages} {279} (\bibinfo {year} {1986})}\BibitemShut {NoStop}%
\bibitem [{\citenamefont {Metzler}\ \emph {et~al.}(2014)\citenamefont
  {Metzler}, \citenamefont {Jeon}, \citenamefont {Cherstvy},\ and\
  \citenamefont {Barkai}}]{metzler2014anomalous}%
  \BibitemOpen
  \bibfield  {author} {\bibinfo {author} {\bibnamefont {Metzler}, \bibfnamefont
  {R.}}, \bibinfo {author} {\bibnamefont {Jeon}, \bibfnamefont {J.-H.}},
  \bibinfo {author} {\bibnamefont {Cherstvy}, \bibfnamefont {A.~G.}}, and\
  \bibinfo {author} {\bibnamefont {Barkai}, \bibfnamefont {E.}},\ }\bibfield
  {title} {\enquote {\bibinfo {title} {Anomalous diffusion models and their
  properties: non-stationarity, non-ergodicity, and ageing at the centenary of
  single particle tracking},}\ }\href@noop {} {\bibfield  {journal} {\bibinfo
  {journal} {Physical Chemistry Chemical Physics}\ }\textbf {\bibinfo {volume}
  {16}},\ \bibinfo {pages} {24128--24164} (\bibinfo {year} {2014})}\BibitemShut
  {NoStop}%
\bibitem [{\citenamefont {Pietronero}(1985)}]{pietronero1985survival}%
  \BibitemOpen
  \bibfield  {author} {\bibinfo {author} {\bibnamefont {Pietronero},
  \bibfnamefont {L.}},\ }\bibfield  {title} {\enquote {\bibinfo {title}
  {Survival probability for kinetic self-avoiding walks},}\ }\href@noop {}
  {\bibfield  {journal} {\bibinfo  {journal} {Physical Review Letters}\
  }\textbf {\bibinfo {volume} {55}},\ \bibinfo {pages} {2025} (\bibinfo {year}
  {1985})}\BibitemShut {NoStop}%
\end{thebibliography}%

\end{document}



\small
\title{Supporting Information for:\\Size-dependent self-avoidance enables superdiffusive migration in macroscopic unicellulars}

\author{Lucas Tr\"oger}
\author{Florian Goirand}
\author{Karen Alim}%
 \email{k.alim@tum.de}
\affiliation{%
 Technical University of Munich,  TUM School of Natural Sciences, Department of Bioscience, Center for Protein Assemblies (CPA),  85748 Garching, Germany
}%

\maketitle

\section{Data analysis}

\subsection{Computation of the MSD exponent and its standard deviation via error propagation}
We compute the instantaneous MSD exponent $\beta$ as the logarithmic derivative of the MSD
\begin{equation*}
    \beta(t) = \frac{\partial\log\mathrm{MSD}(t)}{\partial\log{t}}.
\end{equation*}
This means that the instantaneous exponent can be approximated by
\begin{equation*}
    \beta(t) = \frac{\log\mathrm{MSD}(t+t_\mathrm{min}) - \log\mathrm{MSD}(t)}{\log{(t+t_\mathrm{min})} - \log{(t)}},\quad\mathrm{with}\:t_\mathrm{min}=\SI{4}{min},
\end{equation*}
where $t_\mathrm{min}$ is the acquisition rate of the microscope images.

To compute the standard deviation of $\beta$, we need to propagate the error of the MSD, so its standard deviation. We assume the measurement error of the time to be negligible, which means that we only take into account the standard deviation of the MSD and write $\beta$ as $\beta=c[\log(X)-\log(Y)]$, with $X=\mathrm{MSD}(t+t_\mathrm{min})$, $Y=\mathrm{MSD}(t)$ and $c=1/(\log{(t+t_\mathrm{min})} - \log{(t)})$. This can be approximated by $\beta\approx c*\left[\frac{1}{\langle X\rangle }X - \frac{1}{\langle Y\rangle } Y\right]$. Now, taking the variance on both sides, and using the formula for the variance of a linear combination of variables
\begin{equation*}
    Var(aX+bY) = a^2 Var(X) + b^2 Var(Y) + 2ab*Cov(X,Y),
\end{equation*}
we get
\begin{equation*}
    \frac{\sigma_\beta^2}{c^2} \approx \frac{\sigma_X^2}{\langle X\rangle ^2}+\frac{\sigma_Y^2}{\langle Y\rangle ^2} - \frac{2}{\langle X\rangle \langle Y\rangle }\sigma_{XY}.
\end{equation*}
We have
\begin{equation*}
    \sigma_{XY} = \langle [(X-\langle X\rangle )(Y-\langle Y\rangle )]\rangle  = \langle XY\rangle  -\langle X\rangle \langle Y\rangle ,
\end{equation*}
so 
\begin{equation*}
    \frac{\sigma_\beta^2}{c^2} \approx \frac{\sigma_X^2}{\langle X\rangle ^2}+\frac{\sigma_Y^2}{\langle Y\rangle ^2} - \frac{2\langle XY\rangle }{\langle X\rangle \langle Y\rangle } +2.
\end{equation*}

\subsection{Running/tumbling duration distributions}
\begin{figure}
\centering
\includegraphics[width=\textwidth]{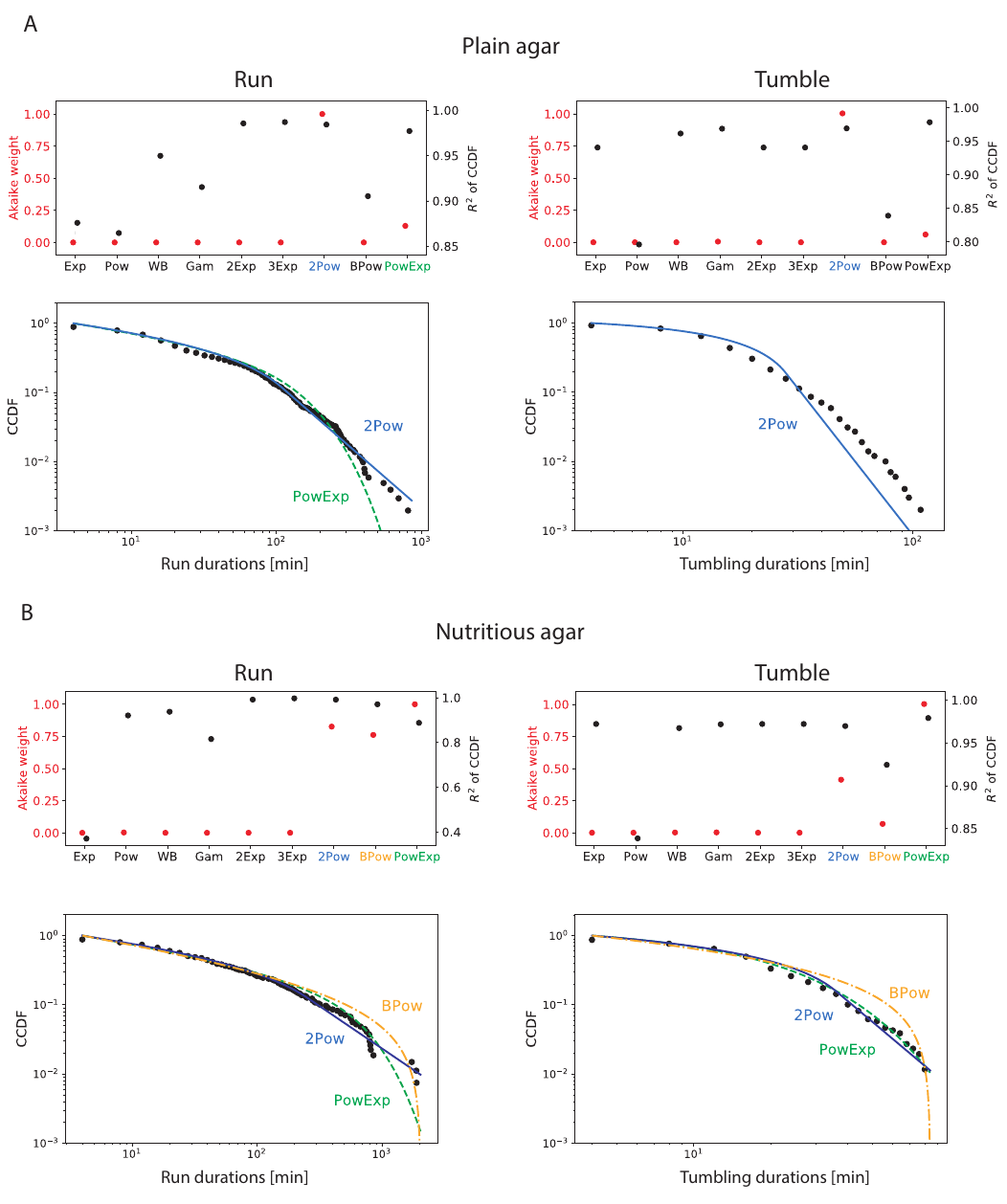}
\caption{Akaike weights of the different tested models and the associated R-square values with respect to the complementary cumulative distribution function (CCDF) of the experimental data for plasmodia migrating on plain agar (A) and nutritious agar (B). The models are Exp=exponential, Pow=power law, WB=weibull, Gam=gamma, 2Exp=two exponentials, 3Exp=three exponentials, 2Pow=two power laws, BPow=bounded power law, PowExp=exponentially truncated power law.}
\label{fig:akaike}
\end{figure}

\begin{figure}
\centering
\includegraphics[width=\textwidth]{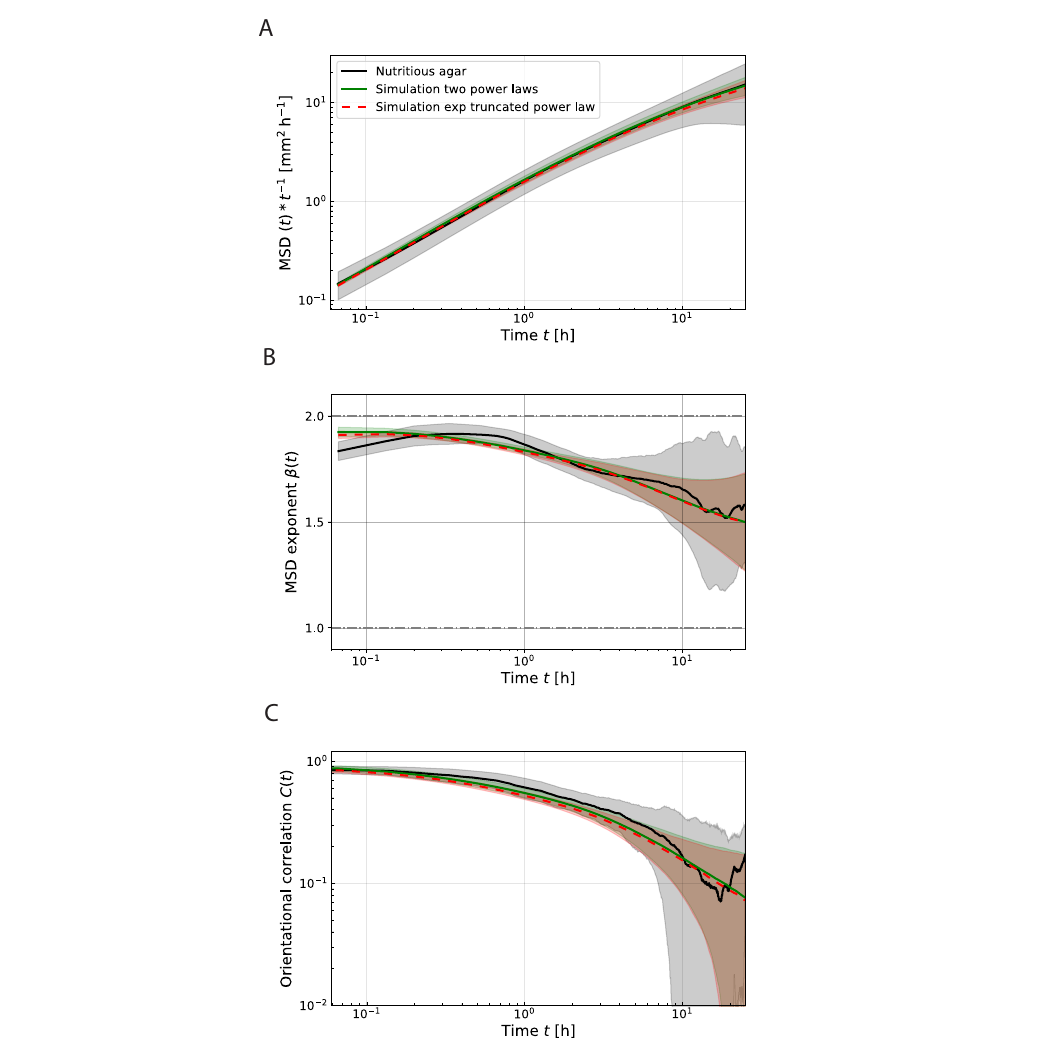}
\caption{Simulation results are the same for two types of run duration distributions: A combination of two power laws (green solid line) and an exponentially truncated power law (dashed red line), compared to the data from nutritious agar (black solid line). (A) Log-log plot of the MSD for the experimental data on nutritious agar and for the simulations using the two different distributions of run durations. (B) Lin-log plot of the MSD exponent $\beta$. (C) Log-log plot of the orientational correlation $C$.}
\label{fig:expcut}
\end{figure}

\begin{figure}
\centering
\includegraphics[width=\textwidth]{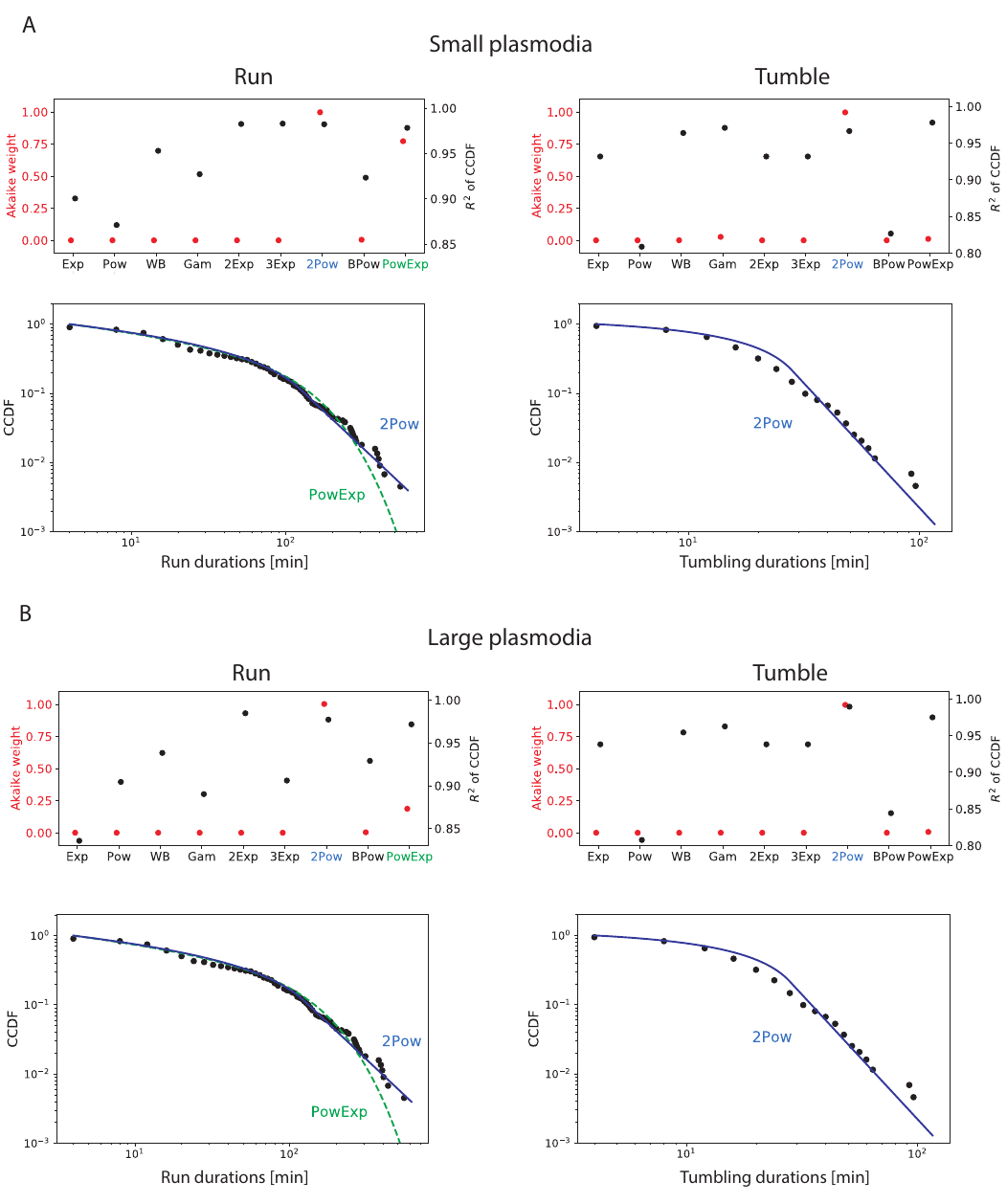}
\caption{Akaike weights of the different tested models and the associated R-square values with respect to the complementary cumulative distribution function (CCDF) of the experimental data for small (A) and large (B) plasmodia. The models are Exp=exponential, Pow=power law, WB=weibull, Gam=gamma, 2Exp=two exponentials, 3Exp=three exponentials, 2Pow=two power laws, BPow=bounded power law, PowExp=exponentially truncated power law.}
\label{fig:akaike2}
\end{figure}

To find the best model describing the distributions of running and tumbling durations, we use maximum likelihood estimation and the Akaike information criterion \citep{anderson2004model}. We find that the best model is a combination of two power laws, except for the plasmodia migrating on nutritious agar, where the best model is an exponentially truncated power law (Figs.~\ref{fig:akaike}, \ref{fig:akaike2} and Table \ref{tab:akaike}). However, simulations show the same results for both types of model (Fig.~\ref{fig:expcut}).

The full PDF of the distribution combining two power laws is 
\begin{equation*}
    \large f(x;\alpha,\chi,b)=A\left[x^{-\alpha}\theta(x-x_\mathrm{min})\theta(b-x) + b^{\chi-\alpha}x^{\chi}\theta(x-b)\right],\quad\mathrm{with}\:x_\mathrm{min}=\SI{4}{min},
\end{equation*}
where
\begin{equation*}
    \large A=1/\left[\frac{1}{1-\alpha}\left(b^{1-\alpha}-x_\mathrm{min}^{1-\alpha}\right)-\frac{b^{1-\alpha}}{1-\chi}\right],
\end{equation*}
and $\theta(x)$ is the Heaviside function. The mean is given by
\begin{equation*}
    \large \langle x\rangle=b\frac{\frac{1}{2-\alpha}\left[1-(x_\mathrm{min}/b)^{2-\alpha}\right] + \frac{1}{\chi-2}}{\frac{1}{1-\alpha}\left[1-(x_\mathrm{min}/b)^{1-\alpha}\right] + \frac{1}{\chi-1}}.
\end{equation*}
The complementary cumulative distribution function used in Figs.~\ref{fig:akaike} and \ref{fig:akaike2} is given by $\bar{F}(x;\alpha,\chi,b)=1-F(x;\alpha,\chi,b)$, where
\begin{equation*}
    \large F(x;\alpha,\chi,b) = A\left\{\frac{1}{1-\alpha}\left(x^{1-\alpha} - x_\mathrm{min}^{1-\alpha}\right)\theta(b-x) + \left[\frac{1}{1-\alpha}\left(b^{1-\alpha} - x_\mathrm{min}^{1-\alpha}\right) + \frac{b^{\beta-\alpha}}{1-\beta}\left(x^{1-\beta} - b^{1-\beta}\right)\right] \theta(x-b)\right\}.
\end{equation*}

\newpage
\section{Simulations}

\subsection{Diffusion constant of a run-and-tumble motion}
We derive the generalized diffusion constant $D$ in terms of the run-and-tumble characteristics under the assumption that the influence of the self-avoidance is negligible. Let a run-and-tumble walk of $N$ steps consist of $N_0$ runs and $N_0$ tumbles. Be $l_{\mathrm{R},i}$ and $l_{\mathrm{T},i}$ the lengths of the $i$th step during a run or tumble, respectively. Further, be $n_\mathrm{R}$ ($n_\mathrm{T}$) the total number of steps spent running (tumbling), $\langle n_\mathrm{R}\rangle $ ($\langle n_\mathrm{T}\rangle $) the average number of steps during a single run (tumble). Then we can write the time-averaged one-step MSD as
\begin{equation*}
    \mathrm{MSD}_\mathrm{1-step}=\frac1N(\Sigma_il_{\mathrm{R},i}^2+\Sigma_kl_{\mathrm{T},k}^2)=\frac{n_\mathrm{R}}{N}\langle {l_\mathrm{R}^2}\rangle +\frac{n_\mathrm{T}}{N}\langle {l_\mathrm{T}^2}\rangle =\frac{N_0}{N}\big(\langle {n_\mathrm{R}}\rangle \langle {l_\mathrm{R}^2}\rangle +\langle {n_\mathrm{T}}\rangle \langle {l_\mathrm{T}^2}\rangle \big) = \frac{N_0}{\tau N}\big(\langle {R}\rangle \langle {l_\mathrm{R}^2}\rangle +\langle {T}\rangle \langle {l_\mathrm{T}^2}\rangle \big),
\end{equation*}
with $\langle{R}\rangle =\tau\langle n_\mathrm{R}\rangle$ and $\langle {T}\rangle =\tau\langle n_\mathrm{T}\rangle$
the average run and tumble durations, and $\tau$ being the duration of one step. With this, we can compute the one-step MSD in terms of the speeds and the generalized diffusion constant $D$: 
\begin{equation*}
    \mathrm{MSD}_\mathrm{1-step}=\frac{N_0}{N}\tau\big(\langle {R}\rangle \langle {v_\mathrm{R}^2}\rangle +\langle {T}\rangle \langle {v_\mathrm{T}^2}\rangle \big) = D\tau^2,
\end{equation*}
with 
\begin{equation*}
    D=\frac{N_0}{\tau N}\big(\langle {R}\rangle \langle {v_\mathrm{R}^2}\rangle +\langle {T}\rangle \langle {v_\mathrm{T}^2}\rangle \big).
\end{equation*}
Note, that with fixed $N$ and $N_0$, $\langle {R}\rangle$ and $\langle {T}\rangle$ are not independent of each other, and that $D$ as a generalized diffusion constant \citep{metzler2014anomalous} has the units of a squared velocity.

\subsection{Parameter sampling}
Running/tumbling durations are drawn from distributions in the form of a combination of two power laws, respectively. For that, we use the inversion method of random variate generation \citep{Devroye1986}. First, we need to find the inverse of the two parts of the CDF individually. For the first part of the combination of power laws, with the CDF
\begin{equation*}
    \large F_1(x;\alpha,b)=\frac{x^{1-\alpha} - x_\mathrm{min}^{1-\alpha}}{(\alpha/b)^{1-\alpha} - x_\mathrm{min}^{1-\alpha}},
\end{equation*}
the inverse is
\begin{equation*}
    \large F_1^{-1}(x;\alpha,b) = \left[x\left(b^{1-\alpha} - x_\mathrm{min}^{1-\alpha}\right) + x_\mathrm{min}^{1-\alpha}\right]^{1/(1-\alpha)}.
\end{equation*}
For the second part, with the CDF
\begin{equation*}
    \large F_2(x;\chi,b)=b^{\chi-1}\left(b^{1-\chi} - x^{1-\chi}\right),
\end{equation*}
it is
\begin{equation*}
    \large F_2^{-1}(x;\chi,b) = b(1-x)^{1/(1-\chi)}.
\end{equation*}
We can now sample a combined power law distribution by first randomly choosing the part of the distribution we want to sample according to its weight and then plugging in for $t$ the respective inverse CDF random variables drawn from a uniform distribution on the interval $[0,1]$.

\subsection{Generating self-avoiding walks}
\begin{figure}
\centering
\includegraphics[width=\textwidth]{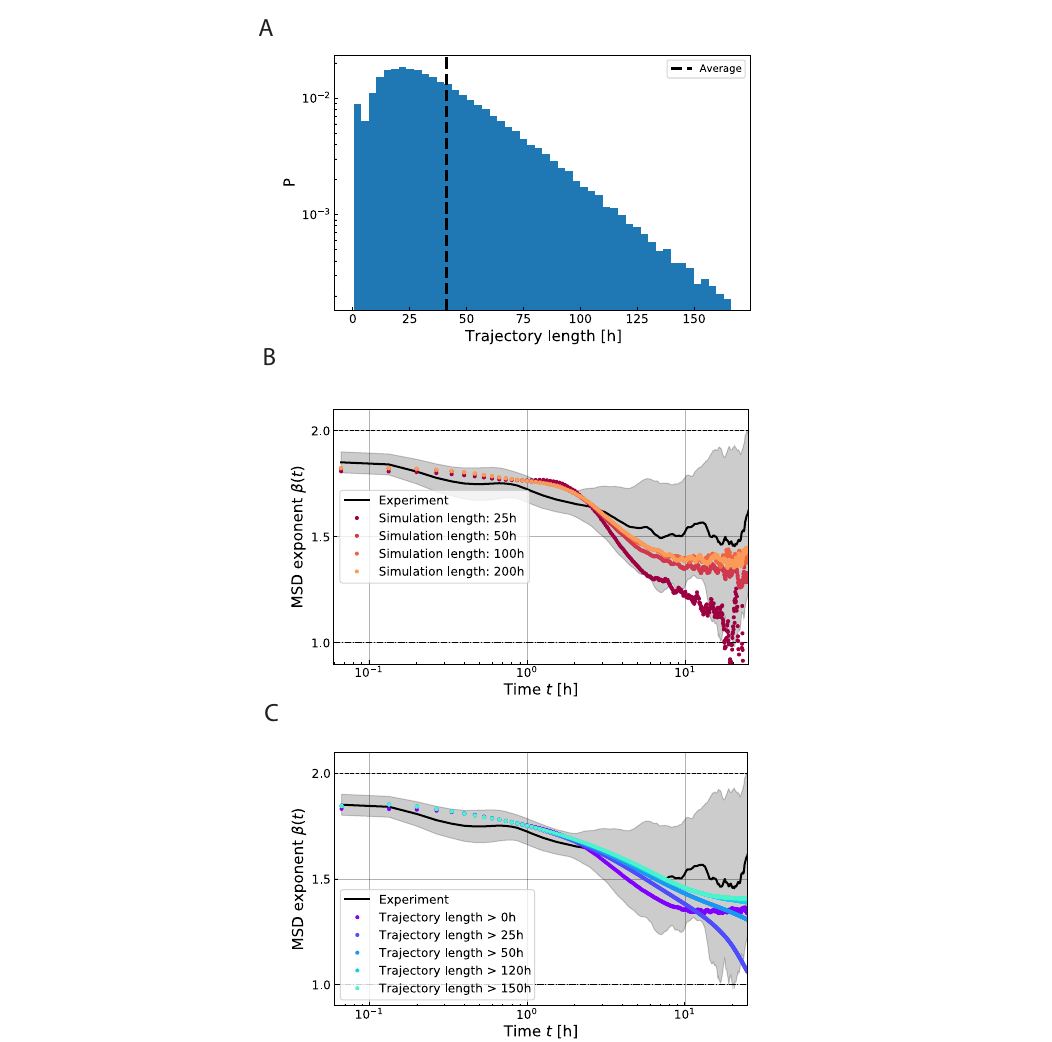}
\caption{Only long trajectories of a length greater than $\SI{120}{h}$ approach the true long time behavior of the MSD exponent $\beta(t)$. (A) Probability distribution of the length of simulated trajectories ($7\times10^5$ realizations of 2500 steps corresponding to 166 hours experimental time). (B) Dependence of $\beta(t)$ on the simulation length ($10^5$ realizations each). (C) Dependence of $\beta(t)$ on the criterion for selecting trajectories. Subsets are formed from the full set of trajectories in (A). The curve displaying the MSD exponent for trajectory lengths $>\SI{25}{\hour}$ exhibits a sampling bias around $\SI{25}{\hour}$ which leads to a drop of the value.}
\label{fig:trapping}
\end{figure}

We construct the run-and-tumble self-avoiding walks in the following way: First, we draw a run duration from the respective power law distribution. Then, for each step of the run phase, we draw a turning angle from the respective exponential distribution and shift the current position by the walking speed times the time step of 4 minutes. If parts of the previous trajectory are encountered, we try a new direction according to the turning angle distribution. Then we draw a tumbling duration from the corresponding power law distribution and perform a diffusive motion with step lengths equal to the tumbling speed times the time step. In this way, walks can enter regions that are completely surrounded by previously visited areas, i.e.~they can get trapped and have a finite length (Fig.~\ref{fig:trapping}). Note, however, that ''only walks much longer than the average walk length before trapping are representative for the true asymptotic behavior`` for this type of self-avoiding walk \citep{hemmer1986trapping,pietronero1985survival}, which is why we only analyze long walks (Fig.~\ref{fig:trapping}C). This type of self-avoiding walk is sometimes called a kinetic, genuine or growing self-avoiding walk \citep{pietronero1985survival,hemmer1986trapping,lyklema1986monte}.

\newpage
\begin{table}[htbp]
  \centering
  \caption{Corrected AIC values of several models computed via maximum-likelihood estimation, and their respective Akaike weights. Minimal AICc (best model) highlighted in red. The models are Exp=exponential, Pow=power law, WB=weibull, Gam=gamma, 2Exp=two exponentials, 3Exp=three exponentials, 2Pow=two power laws, BPow=bounded power law, PowExp=exponentially truncated power law.}
  \resizebox{\columnwidth}{!}{%
    \begin{tabular}{p{15em}cccccccccc}
    \multicolumn{4}{r}{} &
    \multicolumn{6}{p{30em}}{}            &  \\
    \multicolumn{1}{r}{} & \multicolumn{1}{p{3em}}{\textbf{Exp}} & \multicolumn{1}{p{3em}}{\textbf{Pow}} & \multicolumn{1}{p{3em}}{\textbf{WB}} & \multicolumn{1}{p{3em}}{\textbf{Gam}} & \multicolumn{1}{p{3em}}{\textbf{2Exp}} & \multicolumn{1}{p{3em}}{\textbf{3Exp}} & \multicolumn{1}{p{3em}}{\textbf{2Pow}} & \multicolumn{1}{p{3em}}{\textbf{BPow}} & \multicolumn{1}{p{3em}}{\textbf{PowExp}} \\
    \rowcolor[rgb]{0.5,  .7,  .9} Plain agar, run durations & \cellcolor[rgb]{ 1,  1,  1} & \cellcolor[rgb]{ 1,  1,  1} & \cellcolor[rgb]{ 1,  1,  1} & \cellcolor[rgb]{ 1,  1,  1} & \cellcolor[rgb]{ 1,  1,  1} & \cellcolor[rgb]{ 1,  1,  1} & \cellcolor[rgb]{ 1,  1,  1} & \cellcolor[rgb]{ 1,  1,  1} & \cellcolor[rgb]{ 1,  1,  1} & \cellcolor[rgb]{ 1,  1,  1} \\
    AICc  & 7171.30 & 6947.80 & 7103.40 & 7150.00 & 6992.30 & 6996.30 & \textcolor[rgb]{ 1,  0,  0}{6669.60} & 6789.70 & 6688.10  \\
    Delta & 501.70 & 278.20 & 433.80 & 480.40 & 322.70 & 326.70 & 0.00  & 120.10 & 18.50 &  \\
    Akaike weight & 6.17E-25 & 3.76E-14 & 1.45E-19 & 1.37E-21 & 9.67E-15 & 6.48E-15 & 1.00E+00 & 6.08E-06 & 1.57E-01 &  \\
    \rowcolor[rgb]{0.5,  .7,  .9} Plain agar, tumbling durations & \cellcolor[rgb]{ 1,  1,  1} & \cellcolor[rgb]{ 1,  1,  1} & \cellcolor[rgb]{ 1,  1,  1} & \cellcolor[rgb]{ 1,  1,  1} & \cellcolor[rgb]{ 1,  1,  1} & \cellcolor[rgb]{ 1,  1,  1} & \cellcolor[rgb]{ 1,  1,  1} & \cellcolor[rgb]{ 1,  1,  1} & \cellcolor[rgb]{ 1,  1,  1} \\
    AICc  & 5234.90 & 5538.70 & 4902.50 & 4811.60 & 5238.90 & 5242.90 & \textcolor[rgb]{ 1,  0,  0}{4764.70} & 5483.60 & 4790.00 \\
    Delta & 470.20 & 774.00 & 137.80 & 46.90 & 474.20 & 478.20 & 0.00  & 718.90 & 25.30 &  \\
    Akaike weight & 3.80E-21 & 2.43E-34 & 1.04E-06 & 9.19E-03 & 2.55E-21 & 1.71E-21 & 1.00E+00 & 6.01E-32 & 7.97E-02 &  \\
    \rowcolor[rgb]{ 1,  0.4,  1} Nutritious agar, run durations & \cellcolor[rgb]{ 1,  1,  1} & \cellcolor[rgb]{ 1,  1,  1} & \cellcolor[rgb]{ 1,  1,  1} & \cellcolor[rgb]{ 1,  1,  1} & \cellcolor[rgb]{ 1,  1,  1} & \cellcolor[rgb]{ 1,  1,  1} & \cellcolor[rgb]{ 1,  1,  1} & \cellcolor[rgb]{ 1,  1,  1} & \cellcolor[rgb]{ 1,  1,  1} & \cellcolor[rgb]{ 1,  1,  1} \\
    AICc  & 2410.40 & 2193.40 & 2270.50 & 2309.90 & 2223.50 & 2221.40 & 2140.20 & 2141.00 & \textcolor[rgb]{ 1,  0,  0}{2138.60}  \\
    Delta & 271.80 & 54.80 & 131.90 & 171.30 & 84.90 & 82.80 & 1.60  & 2.40  & 0.00  &  \\
    Akaike weight & 1.57E-12 & 4.17E-03 & 1.87E-06 & 3.64E-08 & 2.06E-04 & 2.54E-04 & 8.52E-01 & 7.87E-01 & 1.00E+00 &  \\
    \rowcolor[rgb]{ 1,  0.4,  1} Nutritious agar, tumble durations & \cellcolor[rgb]{ 1,  1,  1} & \cellcolor[rgb]{ 1,  1,  1} & \cellcolor[rgb]{ 1,  1,  1} & \cellcolor[rgb]{ 1,  1,  1} & \cellcolor[rgb]{ 1,  1,  1} & \cellcolor[rgb]{ 1,  1,  1} & \cellcolor[rgb]{ 1,  1,  1} & \cellcolor[rgb]{ 1,  1,  1} & \cellcolor[rgb]{ 1,  1,  1} & \cellcolor[rgb]{ 1,  1,  1} \\
    AICc  & 1381.40 & 1413.10 & 1335.80 & 1325.60 & 1385.50 & 1389.60 & 1279.20 & 1295.30 & \textcolor[rgb]{ 1,  0,  0}{1271.20}  \\
    Delta & 110.20 & 141.90 & 64.60 & 54.40 & 114.30 & 118.40 & 8.00  & 24.10 & 0.00  &  \\
    Akaike weight & 1.64E-05 & 6.88E-07 & 1.56E-03 & 4.34E-03 & 1.09E-05 & 7.21E-06 & 4.49E-01 & 8.98E-02 & 1.00E+00 &  \\
    \rowcolor[rgb]{ .973,  .569,  .514} Small plasmodia, run durations & \cellcolor[rgb]{ 1,  1,  1} & \cellcolor[rgb]{ 1,  1,  1} & \cellcolor[rgb]{ 1,  1,  1} & \cellcolor[rgb]{ 1,  1,  1} & \cellcolor[rgb]{ 1,  1,  1} & \cellcolor[rgb]{ 1,  1,  1} & \cellcolor[rgb]{ 1,  1,  1} & \cellcolor[rgb]{ 1,  1,  1} & \cellcolor[rgb]{ 1,  1,  1} & \cellcolor[rgb]{ 1,  1,  1} \\
    AICc  & 3968.60 & 3755.50 & 3917.30 & 3952.20 & 3838.20 & 3839.40 & \textcolor[rgb]{ 1,  0,  0}{3632.10} & 3683.20 & 3647.20 \\
    Delta & 336.50 & 123.40 & 285.20 & 320.10 & 206.10 & 207.30 & 0.00  & 51.10 & 15.10 &  \\
    Akaike weight & 2.43E-15 & 4.37E-06 & 4.11E-13 & 1.25E-14 & 1.12E-09 & 9.93E-10 & 1.00E+00 & 6.04E-03 & 2.21E-01 &  \\
    \rowcolor[rgb]{ .973,  .569,  .514} Small plasmodia, tumble durations & \cellcolor[rgb]{ 1,  1,  1} & \cellcolor[rgb]{ 1,  1,  1} & \cellcolor[rgb]{ 1,  1,  1} & \cellcolor[rgb]{ 1,  1,  1} & \cellcolor[rgb]{ 1,  1,  1} & \cellcolor[rgb]{ 1,  1,  1} & \cellcolor[rgb]{ 1,  1,  1} & \cellcolor[rgb]{ 1,  1,  1} & \cellcolor[rgb]{ 1,  1,  1} & \cellcolor[rgb]{ 1,  1,  1} \\
    AICc  & 2957.10 & 3112.70 & 2782.20 & 2735.20 & 2961.10 & 2965.30 & \textcolor[rgb]{ 1,  0,  0}{2656.30} & 2908.80 & 2706.40  \\
    Delta & 300.80 & 456.40 & 125.90 & 78.90 & 304.80 & 309.00 & 0.00  & 252.50 & 50.10 &  \\
    Akaike weight & 8.64E-14 & 1.51E-20 & 3.41E-06 & 3.74E-04 & 5.79E-14 & 3.80E-14 & 1.00E+00 & 1.08E-11 & 6.67E-03 &  \\
    \rowcolor[rgb]{ .73,  .816,  .64} Large plasmodia, run durations & \cellcolor[rgb]{ 1,  1,  1} & \cellcolor[rgb]{ 1,  1,  1} & \cellcolor[rgb]{ 1,  1,  1} & \cellcolor[rgb]{ 1,  1,  1} & \cellcolor[rgb]{ 1,  1,  1} & \cellcolor[rgb]{ 1,  1,  1} & \cellcolor[rgb]{ 1,  1,  1} & \cellcolor[rgb]{ 1,  1,  1} & \cellcolor[rgb]{ 1,  1,  1} & \cellcolor[rgb]{ 1,  1,  1} \\
    AICc  & 3199.90 & 3191.20 & 3183.90 & 3196.70 & 3150.80 & 3154.50 & \textcolor[rgb]{ 1,  0,  0}{3033.30} & 3082.00 & 3035.60 \\
    Delta & 166.60 & 157.90 & 150.60 & 163.40 & 117.50 & 121.20 & 0.00  & 48.70 & 2.30  &  \\
    Akaike weight & 5.82E-08 & 1.39E-07 & 2.88E-07 & 8.01E-08 & 7.89E-06 & 5.45E-06 & 1.00E+00 & 7.67E-03 & 7.95E-01 &  \\
    \rowcolor[rgb]{.73,  .816,  .64} Large plasmodia, tumble durations & \cellcolor[rgb]{ 1,  1,  1} & \cellcolor[rgb]{ 1,  1,  1} & \cellcolor[rgb]{ 1,  1,  1} & \cellcolor[rgb]{ 1,  1,  1} & \cellcolor[rgb]{ 1,  1,  1} & \cellcolor[rgb]{ 1,  1,  1} & \cellcolor[rgb]{ 1,  1,  1} & \cellcolor[rgb]{ 1,  1,  1} & \cellcolor[rgb]{ 1,  1,  1} & \cellcolor[rgb]{ 1,  1,  1} \\
    AICc  & 2279.80 & 2428.00 & 2123.80 & 2079.20 & 2283.90 & 2288.00 & \textcolor[rgb]{ 1,  0,  0}{2046.60} & 2276.10 & 2087.40 \\
    Delta & 233.20 & 381.40 & 77.20 & 32.60 & 237.30 & 241.40 & 0.00  & 229.50 & 40.80 &  \\
    Akaike weight & 7.45E-11 & 2.73E-17 & 4.44E-04 & 3.84E-02 & 4.95E-11 & 3.28E-11 & 1.00E+00 & 1.08E-10 & 1.69E-02 &  \\
    \end{tabular}%
    }%
  \label{tab:akaike}%
\end{table}%


\begin{figure}
\centering
\includegraphics[width=\textwidth]{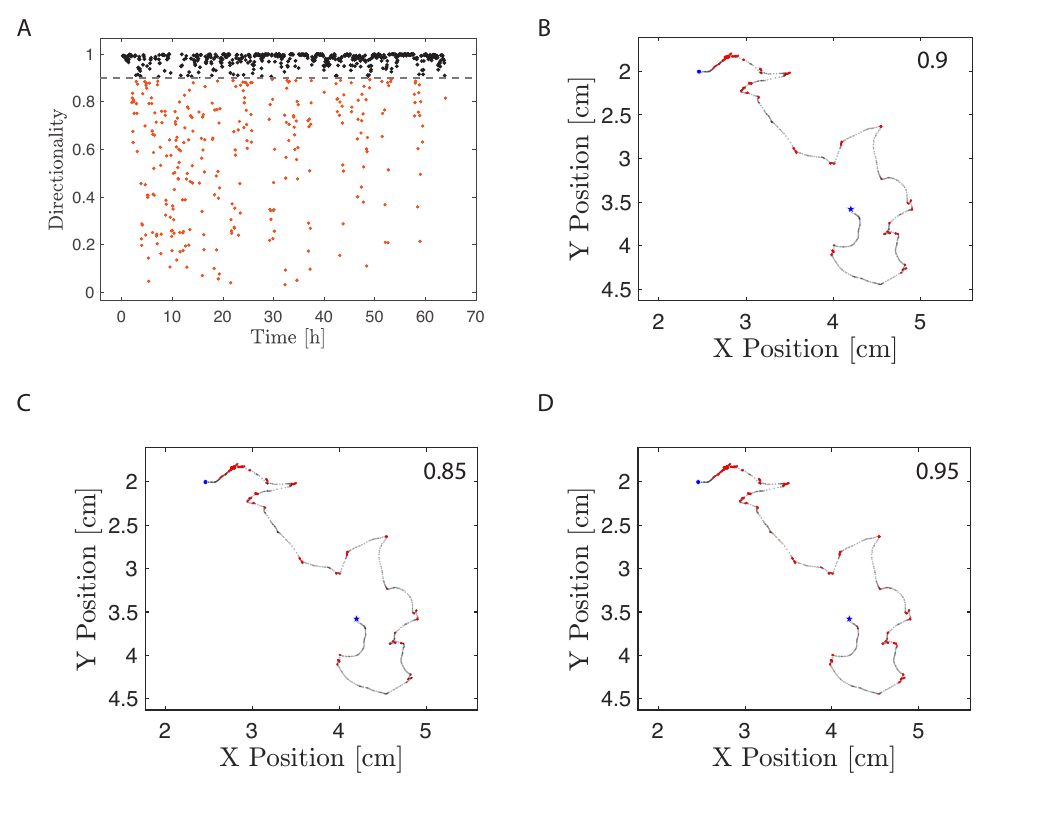}
\caption{Detection of phases of running and tumbling via the directionality of the migration. (A) Directionality of an exemplary plasmodium over the course of its migration. Each scatter point corresponds to a time step, where black points signify a directionality above 0.9, so the running phase, and red dots signify lower directionality, so the tumbling phase. (B) Trajectory of the same plasmodium as in (A), with the same color code of running and tumbling phases. (C-D) Results for directionality thresholds of 0.85 and 0.95.}
\end{figure}

\begin{figure}
\centering
\includegraphics[width=\textwidth]{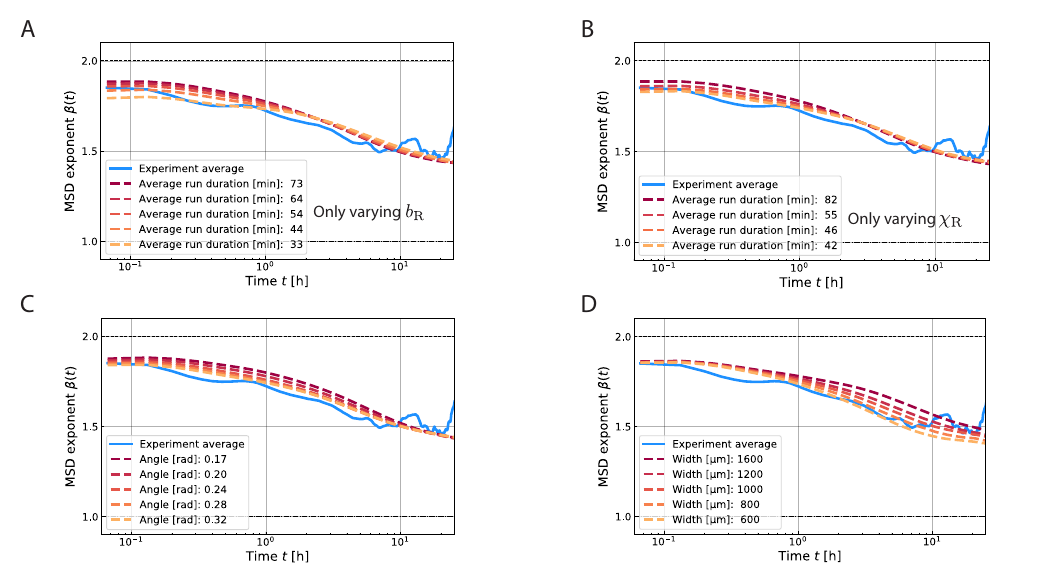}
\caption{Parameter sweeps (5000 trajectories per simulation) reveal the control of the model parameters over the MSD exponent $\beta$. There are ten parameters in the model, four of which are investigated here as examples. One model parameter is varied, while the other parameters are kept constant. Experimental MSD exponent (plain agar) in blue solid line. Simulation MSD exponents in dashed lines. (A) The average run duration $\langle R\rangle$ is varied by changing the power law transition point $b_\mathrm{R}$ in the interval [52, 164] minutes. (B) The average run duration $\langle R\rangle$ is varied by changing the second power law exponent $\chi_\mathrm{R}$ in the interval [2.4, 3.6]. (C) The average turning angle during runs $\phi_\mathrm{R}$ is varied. (D) The median plasmodium width $\bar{w}$ corresponding to the self-avoidance is varied.}
\label{fig:sweeps}
\end{figure}

\begin{figure}
\centering
\includegraphics[width=\textwidth]{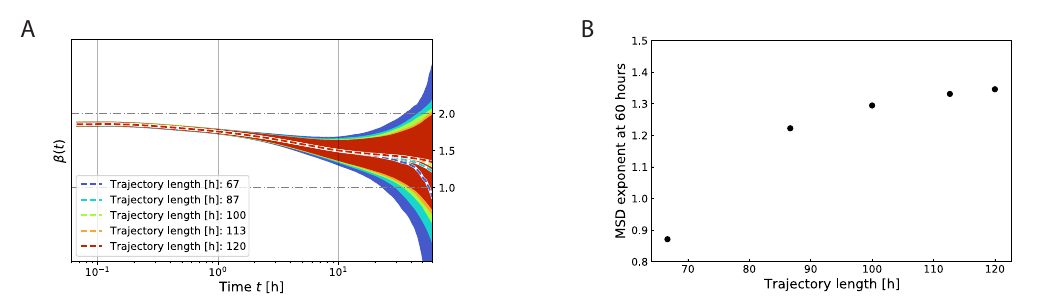}
\caption{Long-time simulations approach an MSD exponent of $\beta=1.5$. Instantaneous MSD exponent $\beta$ depending on the simulated trajectory length for all times (A) and at 60 hours (B).}
\label{fig:exponent}
\end{figure}

\begin{figure}
\centering
\includegraphics[width=\textwidth]{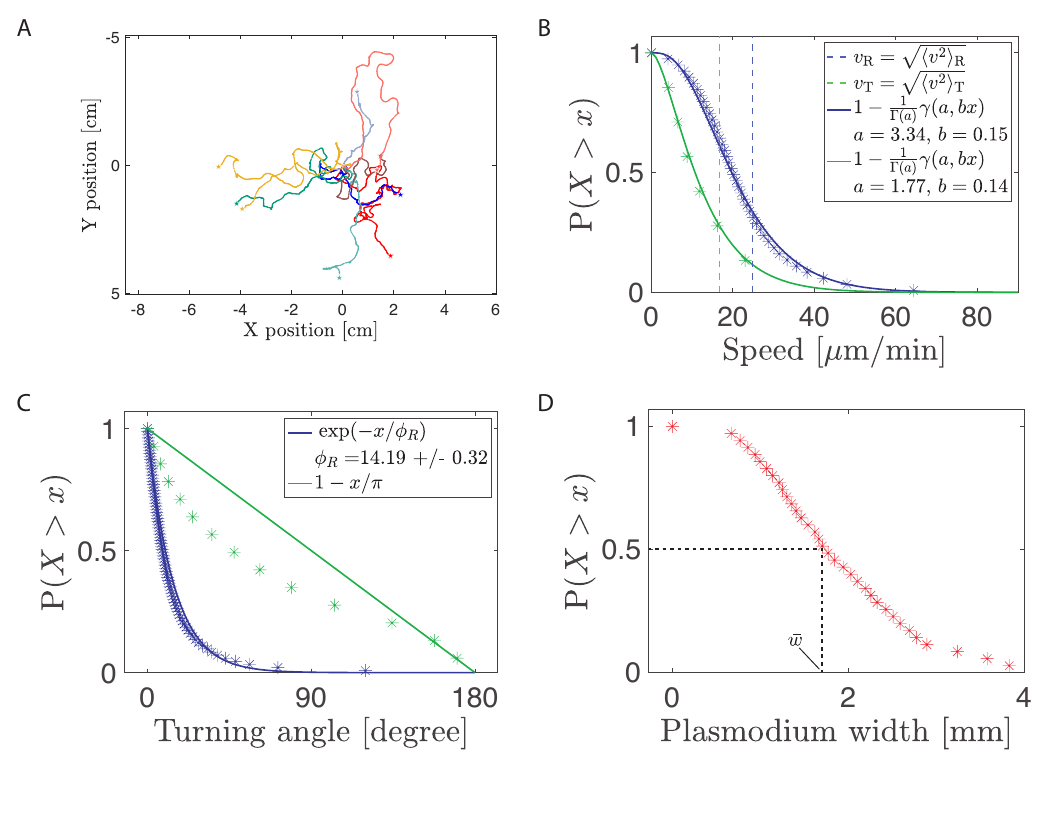}
\caption{Parameter extraction for the migration of plasmodia on nutritious agar. (A) Trajectories of plasmodia migrating on nutritious agar. (B)-(D) Complementary cumulative distribution functions (CCDF), $P(X>x)$, of the analyzed variables, with $X$ denoting the respective variable. Fits in solid lines. (B) Speeds during runs and tumbles with fitted gamma distributions (solid lines). Dashed lines: Root mean squared speeds. (C) Turning angles between consecutive steps during runs and tumbles. Run statistics fitted by an exponential distribution (blue solid line) and tumble statistics approximated by a homogeneous distribution (green solid line). (D) Widths of the plasmodia trails as a measure for the avoided space around the trajectories. The width is estimated as the width of an ellipse fitted to the plasmodium. The dashed line represents the median width of all plasmodia.}
\end{figure}

\begin{figure}
\centering
\includegraphics[width=\textwidth]{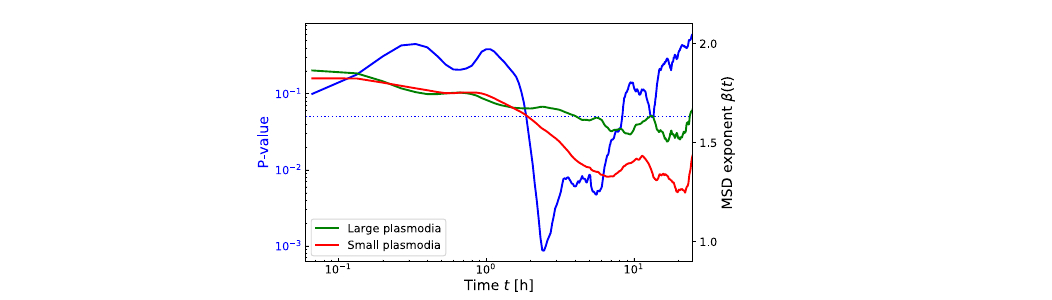}
\caption{Small plasmodia have a lower MSD exponent than large plasmodia on time scales $>\SI{2}{\mathrm{h}}$. Left y-axis shows P-values obtained from a one-tailed Welch's t-test for different times $t$ with the alternative hypothesis that the MSD exponent of small plasmodia is smaller than the MSD exponent of large plasmodia. Choosing a significance level of 5\%, we can conclude that small plasmodia have a lower MSD exponent than large plasmodia on time scales $>\SI{2}{\mathrm{h}}$. For time scales larger than $\SI{10}{\mathrm{h}}$, the experimental values become less and less reliable to make a statement.}
\end{figure}

\begin{figure}
\centering
\includegraphics[width=\textwidth]{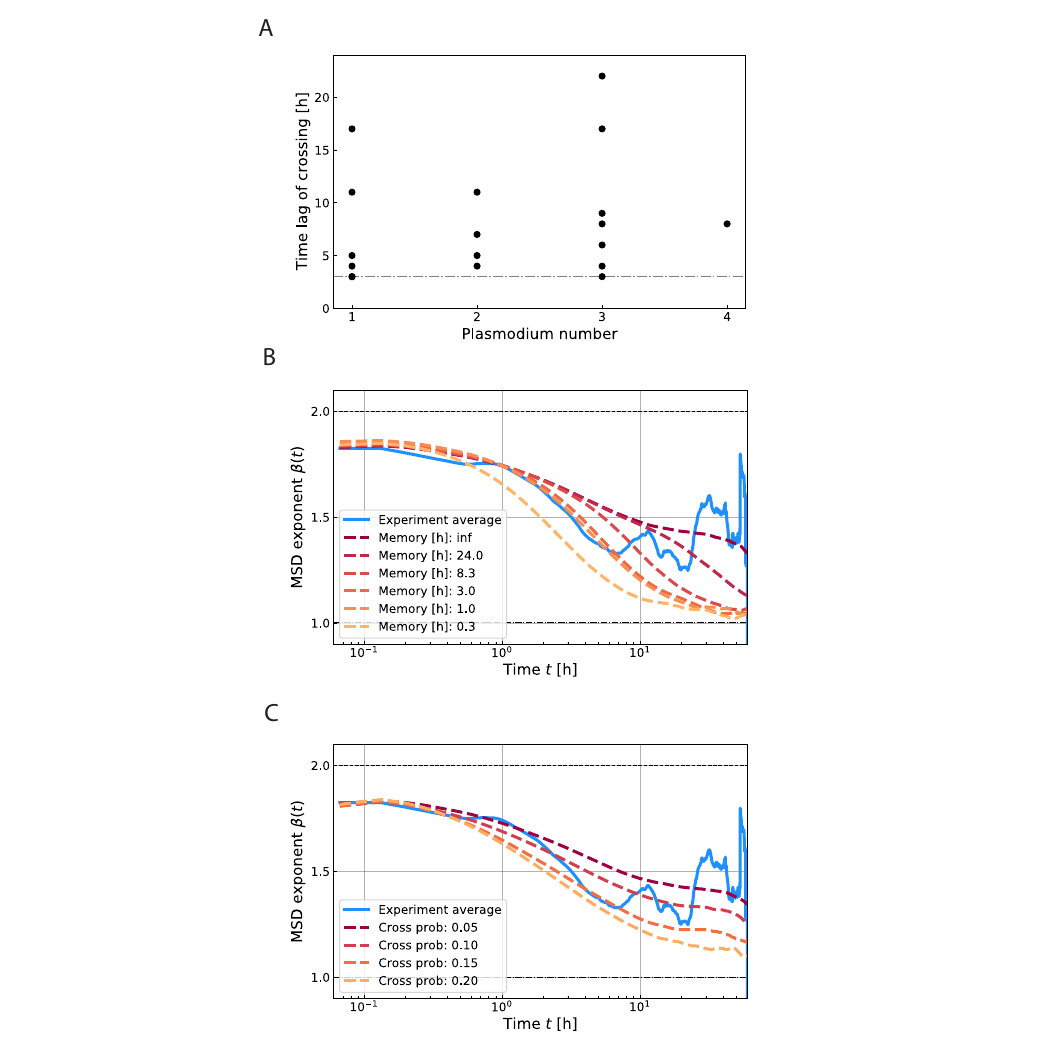}
\caption{Small plasmodia start crossing their own trails after 3 hours. This can well be modeled by a memory-dependent trail. (A) Time lag after which some of the small plasmodia cross their own trail. (B) Changing the memory of the slime trail in simulations (5000 trajectories each) shows that this parameter controls the transition of the MSD exponent to a value of $\beta=1$. Note that an infinite memory corresponds to a strictly self-avoiding walk. (C) Changing the probability to cross the slime trail in simulations (5000 trajectories each) shows that this parameter affects the MSD exponent on all time scales.}
\end{figure}

\begin{figure}
\centering
\includegraphics[width=\textwidth]{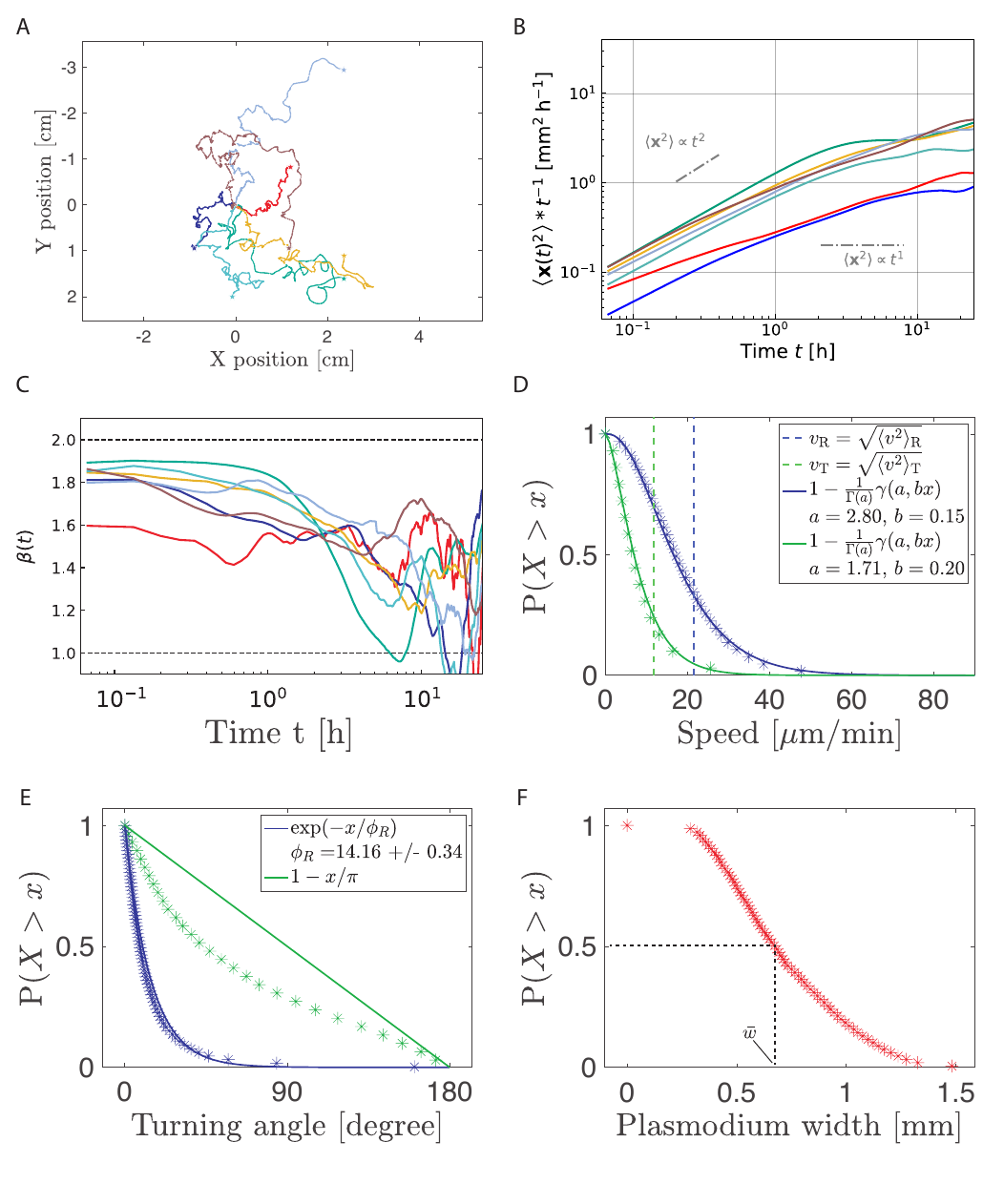}
\caption{Trajectories, MSD and parameter extraction for the small plasmodia group. (A) Trajectories of the small plasmodia shifted to start at the same point. (B) Individual MSDs divided by time of the small plasmodia with colors corresponding to the trajectories in (A). (C) MSD exponents $\beta$ of individual plasmodia. (D)-(F) Complementary cumulative distribution functions (CCDF), $P(X>x)$, of the analyzed variables, with $X$ denoting the respective variable. Fits in solid lines. (D) Speeds during runs and tumbles with fitted gamma distributions (solid lines). Dashed lines: Root mean squared speeds. (E) Turning angles between consecutive steps during runs and tumbles. Run statistics fitted by an exponential distribution (blue solid line) and tumble statistics approximated by a homogeneous distribution (green solid line). (F) Widths of the plasmodia trails as a measure for the avoided space around the trajectories. The width is estimated as the width of an ellipse fitted to the plasmodium. The dashed line represents the median width of all plasmodia.}
\end{figure}

\begin{figure}
\centering
\includegraphics[width=\textwidth]{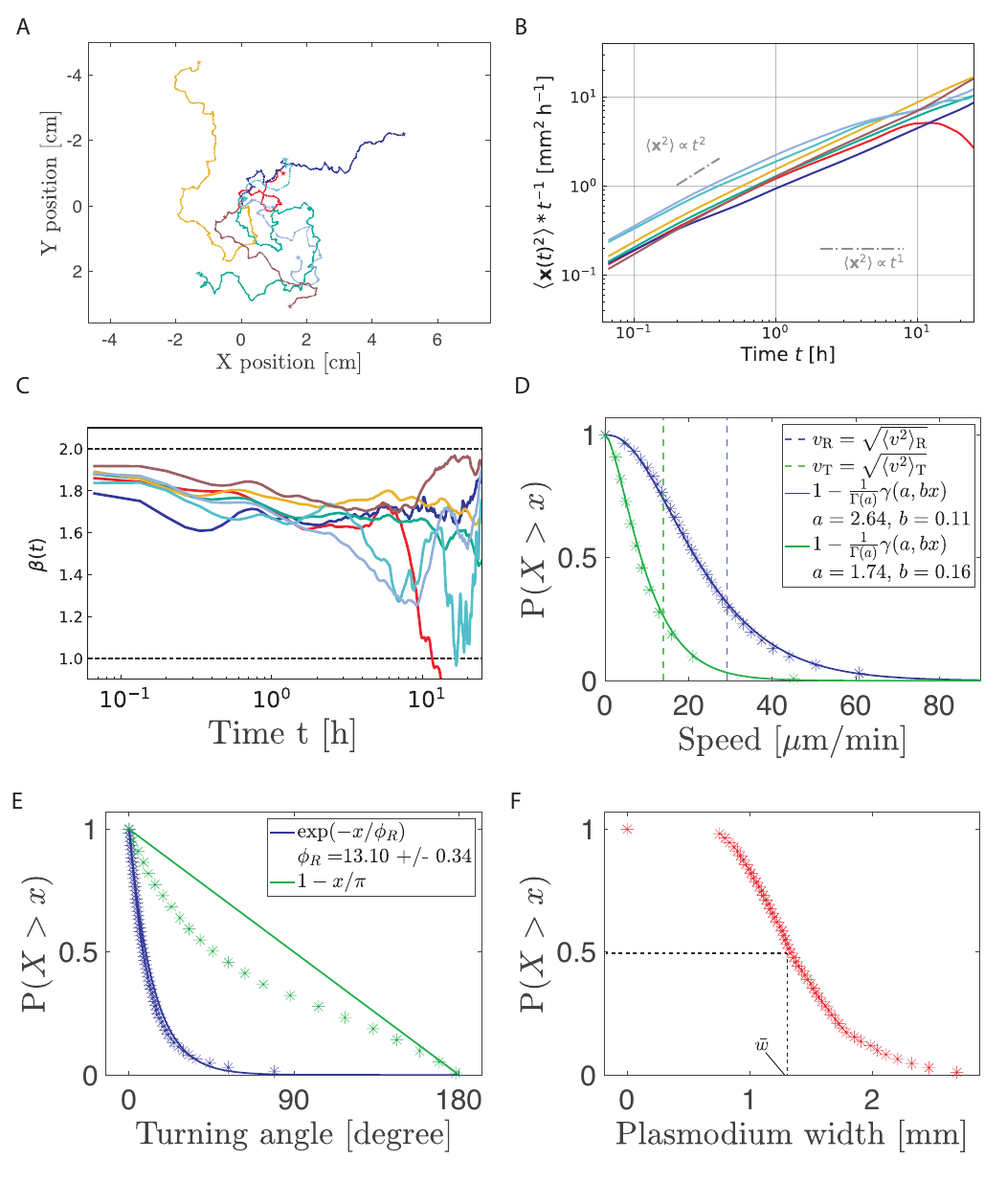}
\caption{Trajectories, MSD and parameter extraction for the large plasmodia group. (A) Trajectories of the large plasmodia shifted to start at the same point. (B) Individual MSDs divided by time of the large plasmodia with colors corresponding to the trajectories in (A). (C) MSD exponents $\beta$ of individual plasmodia. (D)-(F) Complementary cumulative distribution functions (CCDF), $P(X>x)$, of the analyzed variables, with $X$ denoting the respective variable. Fits in solid lines. (D) Speeds during runs and tumbles with fitted gamma distributions (solid lines). Dashed lines: Root mean squared speeds. (E) Turning angles between consecutive steps during runs and tumbles. Run statistics fitted by an exponential distribution (blue solid line) and tumble statistics approximated by a homogeneous distribution (green solid line). (F) Widths of the plasmodia trails as a measure for the avoided space around the trajectories. The width is estimated as the width of an ellipse fitted to the plasmodium as in (A). The dashed line represents the median width of all plasmodia.}
\end{figure}

\begin{figure}
\centering
\includegraphics[width=\textwidth]{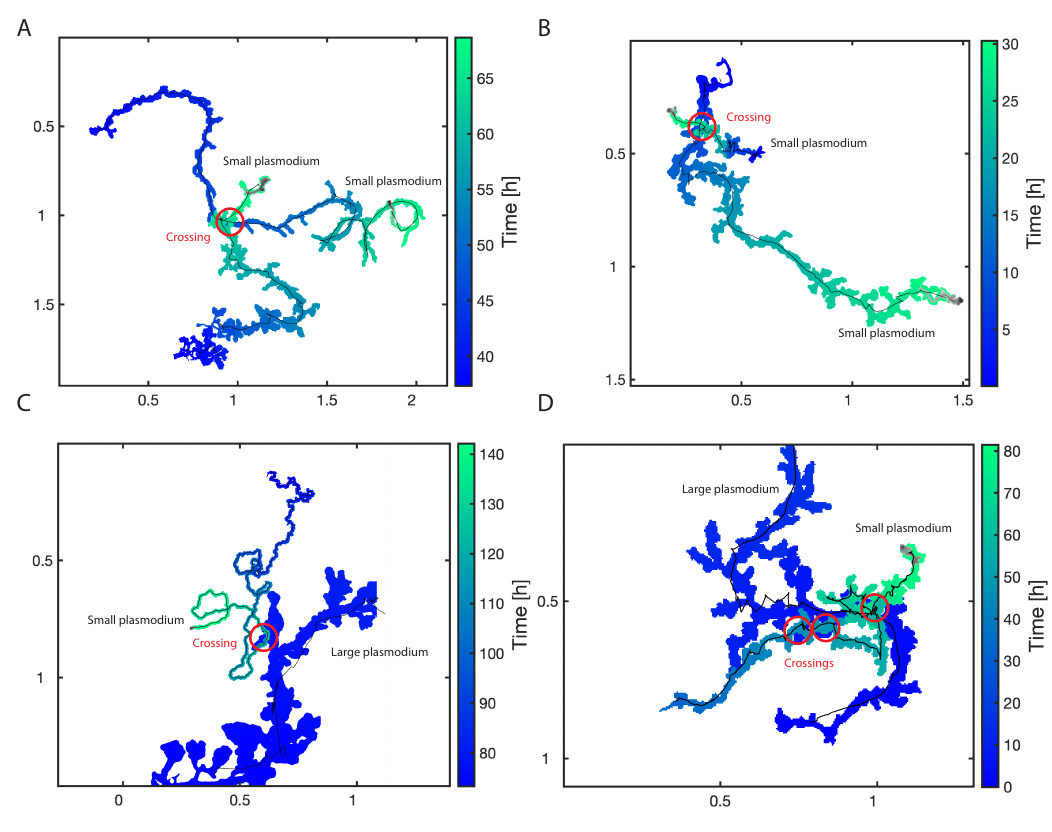}
\caption{Small plasmodia avoid neither the slime trail of small plasmodia nor that of large plasmodia. (A-D) Examples of interaction events between two plasmodia. Plotted are the trails of the individual plasmodia, color coded with the time at which they explore certain areas. These are overlayed by the centroid trajectory (black lines). In case plasmodia are still within the region of interest at the latest time point, greyvalue images are overlayed, marking the position of the plasmodium in the last time frame. Events where plasmodia are crossing the trail of other plasmodia are highlighted by red circles. Ticks on axes in centimeters. (A) Small plasmodium (size: \SI{0.55}{\mm^2}) crossing the slime trail of another small plasmodium (size: \SI{0.28}{\mm^2}). (B) Small plasmodium (size: \SI{0.12}{\mm^2}) crossing the slime trail of another small plasmodium (size: \SI{0.49}{\mm^2}). (C) Small plasmodium (size: \SI{0.03}{\mm^2}) crossing the slime trail of a large plasmodium (size: \SI{2.9}{\mm^2}). (D) Small plasmodium (size: \SI{0.14}{\mm^2}) crossing the slime trail of a large plasmodium (size: \SI{0.77}{\mm^2}).}
\end{figure}

\begin{figure}
\centering
\includegraphics[width=\textwidth]{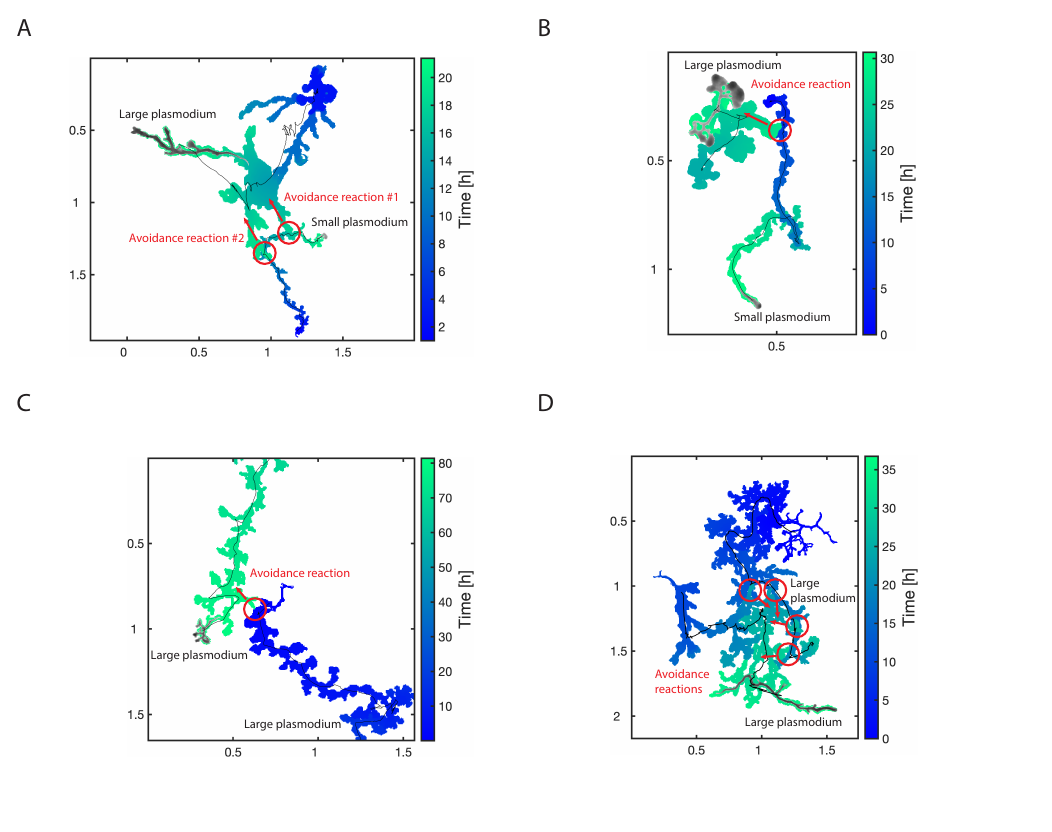}
\caption{Large plasmodia avoid both the slime trail of small plasmodia and that of large plasmodia. (A-D) Examples of interaction events between two plasmodia. Plotted are the trails of the individual plasmodia, color coded with the time at which they explore certain areas. These are overlayed by the centroid trajectory (black lines). In case plasmodia are still within the region of interest at the latest time point, greyvalue images are overlayed, marking the position of the plasmodium in the last time frame. Events where plasmodia recognize the trail of other plasmodia and retract their protrusions as an avoidance reaction are highlighted by red circles. The retracting protrusions are highlighted by red arrows. Ticks on axes in centimeters. (A) Large plasmodium (size: \SI{3.41}{\mm^2}) avoiding the slime trail of a small plasmodium (size: \SI{0.18}{\mm^2}). (B) Large plasmodium (size: \SI{1.40}{\mm^2}) avoiding the slime trail of a small plasmodium (size: \SI{0.24}{\mm^2}). (C) Large plasmodium (size: \SI{0.82}{\mm^2}) avoiding the slime trail of another large plasmodium (size: \SI{0.90}{\mm^2}). (D) Large plasmodium (size: \SI{1.46}{\mm^2}) avoiding the slime trail of another large plasmodium (size: \SI{1.85}{\mm^2}).}
\end{figure}

\begin{figure}
\centering
\includegraphics[width=\textwidth]{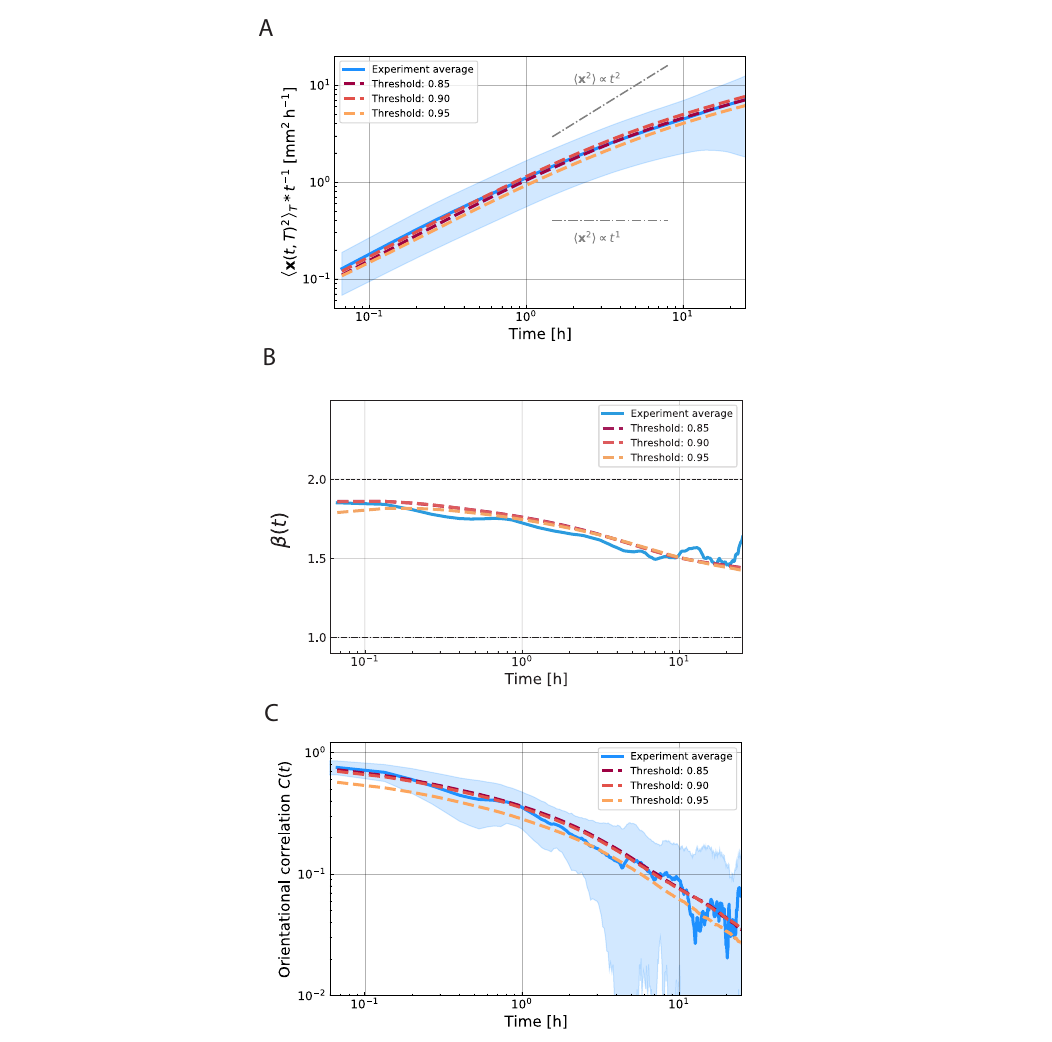}
\caption{Parameter sweep (5000 trajectories per simulation) with respect to the directionality threshold reveals the robustness of the analysis with respect to simulation results, which are performed with model parameters extracted for different values of the directionality threshold (0.85, 0.9 and 0.95) used for the analysis of the run-and-tumble dynamics. Experimental results (plain agar) in blue solid lines. Simulation results in dashed lines. (A) Log-log plot of the MSD divided by time. (B) Lin-log plot of the MSD exponent $\beta$. (C) Log-log plot of the orientational correlation $C$.}
\end{figure}

\clearpage
\bibliography{suppbib}